\documentclass[fleqn,usenatbib]{mnras}

\DeclareRobustCommand{\VAN}[3]{#2}
\let\VANthebibliography\thebibliography
\def\thebibliography{\DeclareRobustCommand{\VAN}[3]{##3}\VANthebibliography}

\usepackage{graphicx}	
\usepackage{amsfonts,amsmath,amssymb,bm,url}	
\usepackage[normalem]{ulem}  
\usepackage[utf8]{inputenc}
\usepackage[font=scriptsize,labelfont=bf]{caption}
\usepackage{caption}
\usepackage{siunitx}
\usepackage{enumitem}
\usepackage{subfig}
\usepackage{lipsum}
\usepackage{xcolor}

\DeclareSIUnit{\erg}{erg}
\DeclareSIUnit{\MeV}{\mega\electronvolt}
\DeclareSIUnit{\fm}{\femto\meter}
\DeclareSIUnit\msol{M_\odot}
\SetLipsumParListSurrounders{\colorlet{oldcolor}{.}\color{gray}}{\color{oldcolor}}

\newcommand{\numberthis}{\addtocounter{equation}{1}\tag{\theequation}}

\makeatletter
\newcommand{\setword}[2]{%
  \phantomsection
  #1\def\@currentlabel{\unexpanded{#1}}\label{#2}%
}
\makeatother

\newcommand{\ls}{\lambda\sigma}

\newcommand{\ep}{\epsilon_k^p}
\newcommand{\en}{\epsilon_{k+q}^n}

\makeatletter

\let\jnl@style=\rm
\def\ref@jnl#1{{\jnl@style#1}}

\def\aj{\ref@jnl{AJ}}                   
\def\actaa{\ref@jnl{Acta Astron.}}      
\def\araa{\ref@jnl{ARA\&A}}             
\def\apj{\ref@jnl{ApJ}}                 
\def\apjl{\ref@jnl{ApJ}}                
\def\apjs{\ref@jnl{ApJS}}               
\def\ao{\ref@jnl{Appl.~Opt.}}           
\def\apss{\ref@jnl{Ap\&SS}}             
\def\aap{\ref@jnl{A\&A}}                
\def\aapr{\ref@jnl{A\&A~Rev.}}          
\def\aaps{\ref@jnl{A\&AS}}              
\def\azh{\ref@jnl{AZh}}                 
\def\baas{\ref@jnl{BAAS}}               
\def\bac{\ref@jnl{Bull. astr. Inst. Czechosl.}}
\def\caa{\ref@jnl{Chinese Astron. Astrophys.}}
\def\cjaa{\ref@jnl{Chinese J. Astron. Astrophys.}}
\def\icarus{\ref@jnl{Icarus}}           
\def\jcap{\ref@jnl{J. Cosmology Astropart. Phys.}}
\def\jrasc{\ref@jnl{JRASC}}             
\def\memras{\ref@jnl{MmRAS}}            
\def\mnras{\ref@jnl{MNRAS}}             
\def\na{\ref@jnl{New A}}                
\def\nar{\ref@jnl{New A Rev.}}          
\def\pra{\ref@jnl{Phys.~Rev.~A}}        
\def\prb{\ref@jnl{Phys.~Rev.~B}}        
\def\prc{\ref@jnl{Phys.~Rev.~C}}        
\def\prd{\ref@jnl{Phys.~Rev.~D}}        
\def\pre{\ref@jnl{Phys.~Rev.~E}}        
\def\prl{\ref@jnl{Phys.~Rev.~Lett.}}    
\def\pasa{\ref@jnl{PASA}}               
\def\pasp{\ref@jnl{PASP}}               
\def\pasj{\ref@jnl{PASJ}}               
\def\rmxaa{\ref@jnl{Rev. Mexicana Astron. Astrofis.}}%
\def\qjras{\ref@jnl{QJRAS}}             
\def\skytel{\ref@jnl{S\&T}}             
\def\solphys{\ref@jnl{Sol.~Phys.}}      
\def\sovast{\ref@jnl{Soviet~Ast.}}      
\def\ssr{\ref@jnl{Space~Sci.~Rev.}}     
\def\zap{\ref@jnl{ZAp}}                 
\def\nat{\ref@jnl{Nature}}              
\def\iaucirc{\ref@jnl{IAU~Circ.}}       
\def\aplett{\ref@jnl{Astrophys.~Lett.}} 
\def\apspr{\ref@jnl{Astrophys.~Space~Phys.~Res.}}
\def\bain{\ref@jnl{Bull.~Astron.~Inst.~Netherlands}}
\def\fcp{\ref@jnl{Fund.~Cosmic~Phys.}}  
\def\gca{\ref@jnl{Geochim.~Cosmochim.~Acta}}   
\def\grl{\ref@jnl{Geophys.~Res.~Lett.}} 
\def\jcp{\ref@jnl{J.~Chem.~Phys.}}      
\def\jgr{\ref@jnl{J.~Geophys.~Res.}}    
\def\jqsrt{\ref@jnl{J.~Quant.~Spec.~Radiat.~Transf.}}
\def\memsai{\ref@jnl{Mem.~Soc.~Astron.~Italiana}}
\def\nphysa{\ref@jnl{Nucl.~Phys.~A}}   
\def\physrep{\ref@jnl{Phys.~Rep.}}   
\def\physscr{\ref@jnl{Phys.~Scr}}   
\def\planss{\ref@jnl{Planet.~Space~Sci.}}   
\def\procspie{\ref@jnl{Proc.~SPIE}}   

\makeatother

\title[]{Proto-neutron star evolution with improved charged-current neutrino-nucleon interactions}

\author[A. Pascal,J. Novak, and M. Oertel]{
  A. Pascal,$^{1}$\thanks{E-mail: aurelien.pascal@obspm.fr}
  J. Novak$^{1}$\thanks{E-mail: jerome.novak@obspm.fr}
  and M. Oertel$^{1}$\thanks{E-mail: micaela.oertel@obspm.fr} \\
$^{1}$Laboratoire Univers et Th\'eories, Observatoire de Paris,
 Universit\'e PSL, CNRS, Universit\'e de Paris, 92190 Meudon, France
}

\date{\today}

\pubyear{2021}

\begin{document}
\label{firstpage}
\pagerange{\pageref{firstpage}--\pageref{lastpage}}
\maketitle

\begin{abstract}
  We perform simulations of the Kelvin-Helmholtz cooling phase of
  proto-neutron stars with a new numerical code in spherical symmetry
  and using the quasi-static approximation. We use for the first time
  the full set of charged-current neutrino-nucleon reactions,
  including neutron decay and modified Urca processes, together with
  the energy-dependent numerical representation for the inclusion of
  nuclear correlations with random-phase approximation. Moreover,
  convective motions are taken into account within the mixing-length
  theory. As we vary the assumptions for computing neutrino-nucleon
  reaction rates, we show that the dominant effect on the cooling
  timescale, neutrino signal and composition of the neutrino-driven
  wind comes from the inclusion of convective motion. Computation of
  nuclear correlations within the random phase approximation, as
  compared to mean field approach, has a relatively small impact.
\end{abstract}

\begin{keywords}
stars:neutron -- neutrinos -- methods: numerical
\end{keywords}


\section{Introduction}
\label{sec:intro}

Neutrons stars are formed during the core-collapse of massive stars at
the end of their evolution. During this collapse the core reaches
supra-nuclear densities and bounces, generating a shock that can
trigger a supernova explosion. The central object left after the
bounce is called a proto-neutron star (PNS). The PNS has a much larger
radius and is still an extremely hot and lepton rich object compared
with cold neutron stars in weak ($\beta$-)equilibrium. During its
evolution, this object will accrete infalling material and slowly
deleptonize, contract and cool down by the Kelvin-Helmholtz mechanism
to become either a stellar black hole or a neutron star
\citep[see e.g.][]{Prakash2001}.

In the last few decades the sensitivity of neutrino detectors has
greatly improved, and a detailed modelling of the emission of
neutrinos associated to core-collapse and subsequent PNS evolution is
needed in order to interpret the data that shall be obtained from the
next galactic supernova \citep[see e.g.][]{Nakazato2021}. This could
lead to a better understanding of the central engine of core-collapse
supernovae (CCSN) and of supernova nucleosynthesis. In addition,
thermodynamic conditions, i.e. temperatures and densities, in PNS are
similar to hypermassive neutron stars formed during mergers of neutron
star binaries \citep[see e.g.][]{Paschalidis2012,iosif-21}, therefore
the study of PNS evolution can lead to a better understanding of
matter properties and neutrino interactions involved in binary
mergers, too.

Long-term multi-dimensional CCSN simulations are, however, still
beyond our reach due to the very different timescales involved:
typical hydrodynamic timescales are orders of magnitude below the
typical deleptonization and cooling time. Therefore we have to make
some compromise to study late-time CCSN and PNS evolution. Work on
this subject exists for more than 25 years. There are two types of
approaches to tackle this problem:
\begin{enumerate}
\item stick with 1D CCSN simulations and either artificially enhance
  the neutrino heating to trigger the explosion \citep[see
  e.g.][]{Fischer2010}, use a low mass O-Ne-Mg core, which are known
  to explode even in 1D simulations \citep[see e.g.][]{Hudepohl2010}, or
  switch from multi-D to 1D by angle averaging after the explosion
  \citep[see e.g.][]{Suwa2014}
\item use the fact that the PNS has a very long evolution timescale
  compared with the neutrino emission and use a quasi-static model of
  its Kelvin-Helmholtz contraction. There are numerous works which
  have used this method, among others we can mention
  \citet{Burrows1986, Keil1995, Sumiyoshi1995, Pons1999} and
  \citet{Roberts2012code}. We have chosen this kind of approach within
  our work.
\end{enumerate}

Although state-of-the-art multi-dimensional simulations are still
bound to about \SI{1}{\second} of evolution, they are very useful for
calibrating simpler models describing evolution at longer
timescales. In particular, they tend to show that the convective
motions in the PNS are extremely significant and change quantitatively
the behaviour of the PNS, see e.g. \citet{Nagakura2020}. A very simple
method to implement convective effects in quasi-static models of PNS
is the use of the mixing length theory (MLT), see
e.g.~\citet{Roberts2012convection} and \citet{Mirizzi2016}. This method
does not cover all features of convection, for instance convective
overshooting cannot be described, but it mimics in a computationally
very efficient way the main effects of convection. We have therefore
adopted the MLT scheme to include convective effects in our work.

Obviously, PNS evolution and the resulting neutrino emission is not
only sensitive to hydrodynamics and matter properties via an equation
of state (EoS), but the interaction of neutrinos with the dense and
hot matter is an extremely important ingredient, too, see e.g. the
discussions in~\citet{Pons1999, Roberts2012neutrinos, MartinezPinedo2012}.
Since the different simulations are computationally expensive, many
authors prefer to employ analytical expressions for the corresponding
interaction rates which imply often very crude approximations. Here,
we will focus the discussion on charged-current neutrino-nucleon
interactions and use standard rates for all other processes. Analytic
expressions for the former can be obtained in limiting cases: assuming
non-interacting nucleons for computing the matrix element and either
neglecting the momentum transfer between the
nucleons~\citep{Bruenn1985} or taking the nucleon momenta on the Fermi
surface~\citep{Yakovlev2001}. The former is valid at low densities
$\lesssim 10^{-4} \mathrm{fm}^{-3}$, conditions typically found in the
early CCSN neutrinosphere, and the latter in degenerate matter,
typically for the cooling of older neutron stars with temperatures
$\ll 1$ MeV. During PNS evolution, the neutrinosphere moves inwards to
higher densities and temperatures vary between a few tens of MeV in
the early times and a few MeV at the later stages. Thus matter in the
relevant regions is neither degenerate nor at sufficiently low
densities and corrections to the above analytical expressions have to
be considered.

In this context, \citet{Roberts2012neutrinos} and
\citet{MartinezPinedo2012} have pointed out the importance of mean
field corrections for nucleon masses and chemical potentials included
in a way consistent with the underlying EoS for
PNS evolution. \citet{Roberts2012neutrinos, Roberts2012convection}
additionally include the full phase space integration
\citep{Reddy1998}, i.e. consider arbitrary momentum transfer between
the nucleons. In \citet{Fischer2020} it is shown that taking into
account inverse neutron decay in addition to the standard electron and
positron capture reactions (and their inverse) influences in
particular the opacity of low energy electron antineutrinos and
consequently the composition of the neutrino-driven wind. In addition,
several authors have pointed out since decades that in dense matter
nuclear correlations beyond mean field can considerably modify the
neutrino opacities: \citet{Burrows:1998cg, Burrows:1998ek, Reddy1999,
  Navarro:1999jn, Margueron:2004sr, Horowitz:2006pj,
  Horowitz:2016gul}. Nuclear correlations treated within the ``Random
Phase Approximation''(RPA) have been considered for PNS evolution by
\citet{Pons1999, Roberts2012convection, Roberts2017handbook} suggesting a
potential influence on the neutrino spectra after several seconds of
evolution when matter starts to become transparent to
neutrinos. However, following \citet{Reddy1999} only grey,
i.e. (anti-)neutrino energy independent, correction factors to the
mean field expressions have been implemented. As discussed in
\citet{Oertel2020}, the importance of RPA correlations is energy
dependent and a density and temperature dependent shift in reaction
thresholds is induced. The full physics can thus not be included into
a grey factor.

Some of these improvements are modifying the analytical rates by
several orders of magnitude, such that it became a priority to perform
CCSN and PNS evolution simulations with state of the art neutrino
interactions in order to test them against previous simulations and
predict the emitted neutrino signal. In \citet{Oertel2020} a first
step has been made in this direction, electron (anti)-neutrino
opacities from charged current neutrino nucleon interactions are
computed including mean field corrections, optionally RPA correlations
and include the full phase space integration. As reactions, electron
and positron capture, neutron and proton decay as well as their
inverse processes have been considered. An interpolation scheme has
been developed which allows to provide the obtained opacity data with
the full dependence on neutrino energy $E_\nu$, baryon number density
$n_B$, temperature $T$ and electron fraction $Y_e$, with tables of
interpolation coefficients publicly available via the \textsc{ComPOSE}
data base~\footnote{\url{https://compose.obspm.fr/}}~\citep{compose}.
The feasibility of CCSN simulations employing these opacities with
only a minor excess in computation time compared with analytical rates
has been demonstrated in \citet{Oertel2020}. In the present work we
perform a step further by using those rates in a simulation of PNS
evolution. Please note that at later times of PNS evolution reactions
can become important which are usually not included in the
simulations. In particular, it should be noted that neutron decay is
one of the main reactions establishing weak equilibrium in cold stars
as well as the so-called modified Urca processes, depending on the
kinematic conditions. Here, we will thus study the impact of RPA
correlations with the full energy dependence and the role of different
reactions for the PNS evolution and the transition to neutrino
transparency. We also provide for the first time simulations with the
most complete set of neutrino charged-current reactions and at the
same time an effective model for convection. This enables us to compare
quantitatively the different effects potentially influencing the PNS evolution.

This paper is organised as follows: in Section~\ref{sec:code} we
present the code and the numerical model used in our simulations and
Section~\ref{sec:cc_int} is focused on the presentation of the various
prescriptions for the computation of charged current reaction
rates. The results of our study are discussed in
Section~\ref{sec:results}, we start with a comparison between models
including convective motions in PNS evolution using MLT and models
without MLT in Section~\ref{ssec:convec} and we then discuss the
impact of the different charged-current neutrino nucleon interactions
in Section~\ref{ssec:cc}. Our conclusions are presented in Section
\ref{sec:summary}.

\section{A code for quasi-static proto-neutron star evolution in spherical symmetry}
\label{sec:code}

This section is devoted to the presentation of the code used in our
PNS simulations. It is based on a quasi-static approach with a
Lagrangian grid and a stationary neutrino transport scheme. Evolution
of entropy and lepton number (Sec.~\ref{ssec:evolution}) is obtained
through neutrino transport (Sec.~\ref{ssec:neutrinos}), and at each
time-step equations for hydrostatic equilibrium are solved
(Sec.~\ref{ssec:hydrostatic}), assuming thus that the hydrodynamics time
scale is much shorter than the neutrino cooling one.

\subsection{Initial models}

Starting with the progenitor model \texttt{s15} from
\citet{Woosley2002}, we run the spherically-symmetric version of the
CCSN code CoCoNuT \citep{Dimmelmeier2005}, using the ``Fast Multigroup
Transport'' (FMT) scheme for neutrino transport \citep{Muller2015},
with the same method as in \citet{Oertel2020}. Each model has been
computed such that the EoS and the neutrino interactions are consistent
with the PNS cooling simulation. To emulate the departure of the shock
and obtain an isolated PNS, we simply discard all the matter behind
the stalled shock, at about $500$~ms after bounce. We thus obtain a
PNS with a baryon mass of $M_B = 1.6 M_\odot$. 

\subsection{Stellar structure}\label{ssec:hydrostatic}

From the quasi-static assumption, gravitational
field is static and therefore we consider a static, spherically-symmetric
spacetime. The metric is written in Schwarzschild gauge, so that we
obtain the usual Tolman–Oppenheimer–Volkoff (TOV)
set of equations for the hydrostatic equilibrium:
\begin{equation}
  ds^2 = - \alpha^2(r) c^2 dt^2 + \psi^2(r) dr^2 + r^2 \left(
    d\theta^2 + \sin^2\theta d\varphi^2 \right)
  \label{eq:metric}
\end{equation}
where $c$ is the speed of light in vacuum, $\alpha$ is called the
lapse function (from 3+1 formalism) and $\psi$ is a radial metric
potential. The system of coordinates is such that $t$ is the time measured
at infinity, $r$ is the areal radius, and $\theta$ and $\varphi$ are
standard angular coordinates. The metric is computed by solving the TOV
equations, assuming that the star is composed of a perfect fluid :
\begin{align}
  \frac{1}{\psi} &= \sqrt{1 - \frac{2Gm}{rc^2} } \\
  \frac{dm}{dr} &= 4\pi r^2 \frac{\mathcal{E}}{c^2} \\
  g(r) &= c^2 \frac{d\ln\alpha}{dr} =  \psi^2  G \left( \frac{m}{r^2}
  + 4\pi r \frac{P}{c^2} \right)~,
  \label{eq:local_g}
\end{align}
where $G$ is the gravitational constant, $m$ is a metric potential
that is equal to the system's ADM mass when taken at the star's
surface, $g$ is the local gravitational acceleration, $\mathcal{E}$
the fluid energy density and $P$ its pressure. The hydrostatic
equilibrium equation is given by
\begin{equation}
  \frac{dP}{dr} = -(\mathcal{E}+P) g(r)~.
\end{equation}
The boundary conditions are given by $m(0) = 0$, $\alpha(R)\psi(R)=1$
and $P(R)=P_s$ where $R$ is the radius of the star and $P_s$ is a
surface pressure, whose value is chosen to have a negligible effect
on the solution while ensuring the numerical stability of the
algorithm. In the simulations presented in this work we took $P_s =
\SI{e-6}{\MeV\per\fm\cubed}$. 

As we consider a Lagrangian evolution scheme, the radial coordinate $r$ is
an unknown, too, and an additional equation is solved together with the previous
system :
\begin{equation}
  \frac{dr}{da} = \frac{1}{4\pi r^2 n_B \psi}~,
\end{equation}
where $a$ is the enclosed baryon number (fixed during the
evolution and used as a Lagrangian coordinate), and $n_B$ is the
baryon number density.

Finally, in order to close the system of equations, we need an EoS
relating thermodynamic variables. In our case it shall be a
3-parameter one, depending namely on temperature $T$, baryon density
$n_B$ and electron fraction $Y_e$. In this work we have used three
different EoS models: (i) the \textit{nuclear statistical equilibrium}
(NSE) model by \citet{Gulminelli2015} (``RG(SLy4)''), employing the
SLy4 effective interaction for nucleons~\citep{Chabanat1997}; (ii) the
NSE model by \citet{Hempel2010}, employing the DD2 effective
interaction for nucleons~\citep{Typel_PRC_2010} (``HS(DD2)''); (iii)
the ``SRO(APR)'' model~\citep{Constantinou2014, Schneider2019}. The
latter is based on the APR EoS~\citep{APR}, which itself is partly
adjusted to the variational calculation of \citet{Akmal_1997}. For all
three models. EoS data have been obtained from the \textsc{CompOSE}
database~\citep{compose}.

\subsection{Evolution equations}\label{ssec:evolution}

The time evolution from one quasi-static configuration (computed using
the TOV system given above) to the next is done by considering the
effects of neutrino interactions on the stellar matter via lepton number
and energy conservation. These are given by
\begin{align}
  \nabla_\mu( u^\mu n_B Y_e ) &= \Gamma_{\bar\nu_e} -
                                \Gamma_{\nu_e} \label{eq:hydro1} \\
  u_\nu \nabla_\mu( T^{\mu\nu} ) &= - (Q_{\nu_e} + Q_{\bar\nu_e} + 4
                                   Q_{\nu_x})~, \label{eq:hydro2}
\end{align}
where $u^\mu$ is the fluid four-velocity, $T^{\mu\nu}$ the fluid's
energy-momentum tensor, $\Gamma_\nu$ the neutrino production rate per
volume unit, and $Q_\nu$ the neutrino heat function for each neutrino
or anti-neutrino flavor $\nu_i$.  For a perfect fluid these equations
can easily be recast as
\begin{align}
    u^\mu \nabla_\mu(Y_e) &= \frac{1}{\alpha } \frac{D Y_e}{D t} =
                            \frac{\Gamma_{\bar\nu_e} -
                            \Gamma_{\nu_e}}{n_B} \label{eq:evol_y_nu}\\
    u^\mu \nabla_\mu (s) &= \frac{1}{\alpha } \frac{Ds}{Dt} =
                           \frac{Q_{\nu_e} + Q_{\bar\nu_e} + 4
                           Q_{\nu_x}}{n_B T}  - \frac{\mu_e
                           (\Gamma_{\bar\nu_e} - \Gamma_{\nu_e}) }{n_B
                           T} \label{eq:evol_s_nu}
\end{align}
where $s$ is the entropy per baryon, $\mu_e$ the electron chemical
potential and $D/Dt$ is the Lagrangian derivative. Both source terms,
$\Gamma_\nu$ and $Q_\nu$, are obtained from a neutrino transport
scheme, by using Eqs. (\ref{eq:nu_src_1}) and (\ref{eq:nu_src_2}), see
the Sec.~\ref{ssec:neutrinos} below.

\subsection{Neutrino transport}\label{ssec:neutrinos}

In order to model neutrino transport, we use the FMT scheme
\citep[see][]{Muller2015}, which relies on a stationary approximation
of the transport equation,
\begin{equation}\label{e:nutransport}
  p^i \frac{\partial f}{\partial x^i} - \Gamma^i_{~\mu\nu} p^\mu p^\nu
  \frac{\partial f}{\partial p^i} = u_\mu p^\mu \mathcal {B}[f]~.
\end{equation}
$f$ denotes here the neutrino distribution function, $p^\mu$ the
neutrino four-momentum and $\mathcal B[f]$ the collision integral
computed in the fluid rest frame.

The solution at high optical depth is obtained with a two-stream
approximation and the solution at low optical depth is obtained with a
two-moment closure. The procedure is detailed in
appendix~\ref{appendix:fmt}, whereas the treatment of the collision
integral is detailed in Section~\ref{sec:cc_int} for charged current
reactions on nucleons and in appendix~\ref{appendix:col_int} for all
other processes. It should be stressed in this context that
we model neutrino-nucleon scattering with full
inelastic rates, as given by~\citet{Thompson2000}.

In this kind of stationary approximation, the source terms are
obtained via the divergence of neutrino fluxes:
\begin{align}
  \Gamma_{\nu} &= \frac{1}{r^2 \psi \alpha} \frac{d}{dr} \left( r^2
                 \alpha F_{\nu, n} \right) \label{eq:nu_src_1} \\
  Q_{\nu} &= \frac{1}{r^2 \psi \alpha^2} \frac{d}{dr} \left( r^2
            \alpha^2 F_{\nu, e} \right) \label{eq:nu_src_2}~,
\end{align}
where $F_{\nu,n}$ is the outgoing number flux of neutrino $\nu$ and
$F_{\nu, e}$ is the outgoing energy flux carried by neutrinos $\nu$.
The total luminosities are then given by
\begin{align*}
  L_{\nu,n} &= 4\pi R^2 \alpha(R) F_{\nu, n}(R) \\
  L_{\nu,e} &= 4\pi R^2 \alpha(R)^2 F_{\nu, e}(R)~,
\end{align*}
where $L_{\nu,n}$ is the neutrino number luminosity (in
$\si{\per \second}$) and $L_{\nu,e}$ is the energy luminosity (in
$\si{\erg\per\second}$).

\subsection{Convection using the mixing length theory}\label{ssec:MLT}

We study the effect of convection on PNS evolution within the MLT,
which models convection in spherical symmetry as a diffusive effect
occurring in zones with unstable stratification. The stability
criterion for the stratified structure of a PNS is given by the Ledoux
criterion \citep[see e.g.][]{Roberts2012convection}:
\begin{equation}
  C_L(r) = \frac{1}{\Gamma_{n_B}} \left( \Gamma_s \frac{\partial \ln
      s}{\partial r} + \Gamma_{Y_e} \frac{\partial \ln Y_e}{\partial
    r} \right) \geq 0~,
  \label{eq:ledoux_criterion_pns}
\end{equation}
where $r$ is the radial coordinate and
$\displaystyle\Gamma_{n_B} = \left( \frac{d\ln P}{d \ln n_B}
\right)_{s,Y_e}$,
$\displaystyle\Gamma_s = \left( \frac{d\ln P}{d \ln s}
\right)_{n_B,Y_e}$,
$\displaystyle\Gamma_{Y_e} = \left( \frac{d\ln P}{d \ln Y_e}
\right)_{n_B,s} $.  \medbreak The quantity $C_L(r) dr$ can be
interpreted as the relative variation of density $\Delta n_B / n_B$
occurring during the small vertical adiabatic displacement of a mass
element over a distance $dr$. Several authors
\citep[e.g.][]{Epstein1979, Keil1996, Roberts2012convection} have
argued that, because the neutrinos are trapped and in equilibrium with
the fluid in large regions of the PNS, one should consider the lepton
fraction instead of the electron fraction to compute the Ledoux
criterion. We think however, that as convection is a purely
hydrodynamic feature appearing when considering multi-dimensional
versions of Eqs.~(\ref{eq:hydro1})--(\ref{eq:hydro2}), one should use
only electron fraction, to be consistent with the hydrodynamic model.

The distinction is then made between
\begin{itemize}[label=\textbullet]
  \item areas where $C_L(r) > 0$, where buoyancy acts as a restoring
    force. They can be subject to \textit{gravity waves} ;
  \item areas where $C_L(r) < 0$, which are unstable and can be
    subject to convective motion ;
  \item areas where $C_L(r) = 0$, which are in a state of
    \textit{neutral buoyancy}. Convective motions tend to bring
    unstable areas towards this state.
\end{itemize}

In the MLT, the convective motion in Ledoux-unstable areas is modelled
with diffusion equations for the entropy and the electron number,
\begin{align}
  \frac{1}{\alpha } \frac{D Y_e}{D t}
  &= \frac{1}{r^2 \alpha \psi} \frac{\partial}{\partial r} \left(
    \alpha r^2 D^{\text{\tiny MLT}} n_B \frac{\partial Y_e}{\partial r}
    \right) \label{eq:evol_y_mlt}\\
  \frac{1}{\alpha } \frac{Ds}{Dt}
  &= \frac{1}{r^2 \alpha \psi} \frac{\partial}{\partial r} \left( \alpha r^2 D^{\text{\tiny MLT}} n_B \frac{\partial s}{\partial r} \right) \label{eq:evol_s_mlt}
\end{align}
where $D^{\text{\tiny MLT}}$ is the MLT diffusion coefficient. We
estimate it by using the same procedure as in \citet{Mirizzi2016}: we
have $D^{\text{\tiny MLT}} = v_{\mathrm{c}}\lambda_P$, where
$v_{\mathrm{c}}$ is the convection velocity and $\lambda_P$ is the
length scale over which convective turnover occurs, or the so-called
\textit{mixing length}. This length scale is assumed to be of the same
order of magnitude as that of pressure variation,
\begin{equation}
  \lambda_P = \xi  \left( \frac{\partial \ln P}{\partial r} \right)^{-1}
\end{equation}
where $\xi$ is a coefficient of order unity. The exact value of $\xi$
does not have a strong influence on the results, and we use $\xi=1$ as
the standard value. We have checked that varying its value by $\pm
20\%$ does not change the results of our simulations within the
overall accuracy of the model.

The convection velocity $v_\mathrm{c}$ is estimated using energy
conservation during a vertical displacement of $\lambda_P$:
\begin{equation}
  v_{\mathrm{c}} = \begin{cases}
    \lambda_P \sqrt{2 g |C_L|} & \text{if } C_L(r) \leq 0 \\
    0 & \text{if } C_L(r) \geq 0~,
\end{cases}
\end{equation}
where $g$ is the local gravitational acceleration introduced in
Eq.~(\ref{eq:local_g}).

The PNS evolution equations are then solved with a semi-implicit
scheme: the neutrino part (namely Eqs.~(\ref{eq:evol_y_nu}) and
(\ref{eq:evol_s_nu})) is solved with an explicit scheme whereas the
MLT part (Eqs. (\ref{eq:evol_y_mlt}) and (\ref{eq:evol_s_mlt})) is
solved implicitly. The timestep is limited by the relative change in
$s$ and $Y_e$ induced by the neutrino sources. The explicit
integration scheme for the neutrino part has the advantage of allowing
to perform larger and larger timesteps as the neutrino emissivity
decreases while not having to deal with the heavy computational cost
of an implicit scheme for neutrino transport.

\section{Treatment of charged current interactions with nucleons}
\label{sec:cc_int}

In this section we present the treatment of charged current
neutrino-nucleon interactions. For such processes the collision
integral in the neutrino transport equation~\eqref{e:nutransport} is
linear in $f_\nu$ and can be written in the form:
\begin{equation}
  \mathcal{B}[f] = j(1-f) - \frac{1}{\lambda} f = \kappa^* (f^{(eq)} - f)~,
\end{equation}
where $j$ is the emissivity, $\lambda$ the absorption mean free path
and $\kappa^* = j + \frac{1}{\lambda}$ is the opacity corrected for
stimulated absorption.  It should be stressed that in our work we
consider only charged-current processes involving electrons, and we
neglect the effect of muonic processes. Such processes have long been
thought to have a negligible influence on CCSN because of the
relatively low abundance of muons compared with electrons due to their
much higher mass. The recent work of \citet{Bollig2017} has, however,
shown that the appearance of muons has a significant effect on the
CCSN evolution by softening the EoS and muonic processes have gained
interest. Thus, the recent study by \citet{Fischer:2020vie}
has quantified the effect of muonic charged-current processes on the
neutrino luminosities. We will consider the impact of muonic processes
on PNS evolution in a future work.

\begin{figure*}
  \centering
  \subfloat[$T = \SI{2}{\MeV}$]{\includegraphics[width=0.45\textwidth]{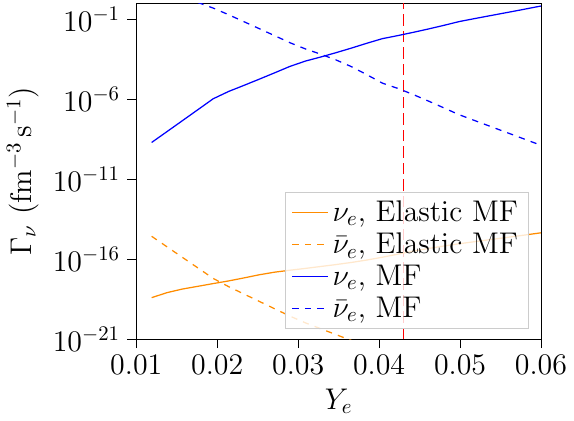}}\hfill
  \subfloat[$T = \SI{0.3}{\MeV}$]{\includegraphics[width=0.45\textwidth]{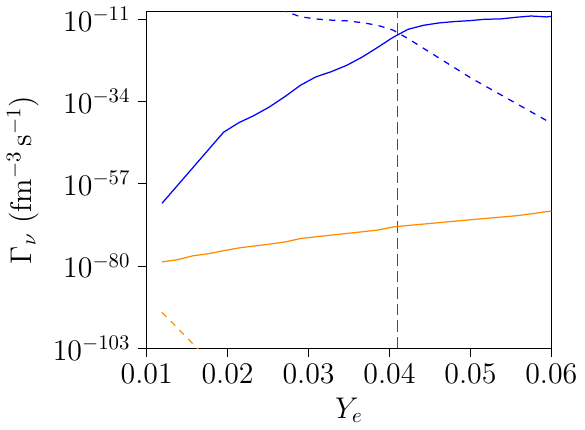}}
  \caption{Electron neutrino and antineutrino production rates per
    unit volume, at the density
    $n_B=\SI{0.1}{\per\femto\meter\cubed}$, for two different
    temperatures as functions of the electron fraction. Both
    temperature values are low enough for matter to be considered as
    transparent. The RG(SLy4) EoS has been used.  $\beta$-equilibrium
    is obtained if neutrino and antineutrino rates are equal, i.e. at
    the intersections of the $\nu_e$ and $\bar\nu_e$ curves.  The
    vertical red dashed line represents the equilibrium electron
    fraction predicted by the Fermi surface approximation.}
  \label{fig:beta_eq_elastic}
\end{figure*}

\subsection{Direct Urca processes}

Direct Urca processes are the dominant and simplest charged current
neutrino-nucleon interactions involving electron neutrinos and
antineutrinos. The processes for electron neutrinos are

\begin{equation}
  p + e^- \leftrightarrows n + \nu_e \qquad \qquad p \leftrightarrows n + e^+ + \nu_e~,
\label{eq:durca}
\end{equation}

and the corresponding processes for antineutrinos are

\begin{equation}
  n + e^+ \leftrightarrows p + \bar\nu_e \qquad\qquad n
  \leftrightarrows p + e^- + \bar\nu_e~.
\end{equation}

In this work we employ three different approximations for computing
the rates for the above processes. The first one is the
\textit{elastic approximation} (``elastic MF'') in which the momentum
transfer between the nucleons is neglected
\citep[see][]{Bruenn1985}. Mean field effects are added as effective
masses and single particle potentials for the
nucleons~\citep{Reddy1998}. In addition we use the \textit{Mean Field}
(``MF'') approximation with full phase space integration, and a scheme
including nuclear correlations within the \textit{Random Phase
  Approximation (RPA)}. In particular, we used the so-called
``RPA~$t_3'$'' approximation, which considers an additional repulsive
term in the residual interaction in the axial channel to prevent
instabilities at high densities. For details about the different
approximation schemes and the practical implementation, we refer to
\citet{Oertel2020}. In particular, ``MF'' and ``RPA~$t_3'$'' rates are
provided via tabulated interpolation coefficients for fully energy
dependent precomputed opacities in order to avoid the heavy additional
computational cost of the integration of opacities with full space or
in RPA.

The elastic approximation is still very commonly used in many recent
simulations of PNS evolution \citep[see e.g.][]{Li:2020ujl,
  Nakazato:2020ogl}, as it is quite accurate at high temperatures
and neutrino energies and relatively low densities, i.e. for the
typical conditions close to the neutrinosphere in CCSN. We should,
however, point out again that the elastic approximation looses
accuracy during PNS evolution with the neutrinosphere moving to higher
densities and the subsequent cooling of the PNS. At higher densities,
the elastic rates can differ from the full ones by orders of magnitude
and in addition, $\beta$-equilibrium is displaced towards more neutron
rich conditions. Both effects are illustrated in
Fig.~\ref{fig:beta_eq_elastic}, where electron neutrino and
antineutrino production rates are shown as functions of the electron
fraction for two different temperatures and a baryon number density of
$n_B = 0.1 \mathrm{fm}^{-3}$, employing either the elastic
approximation or the full phase space integration, including mean
field effects in both cases. For the shown conditions, matter can be
considered as transparent to (anti)-neutrinos and $\beta$-equilibrium
is obtained if neutrino and antineutrino rates are equal, i.e. at the
intersections of the $\nu_e$ and $\bar\nu_e$ curves. It is obvious
that with the elastic rates, $\beta$-equilibrium occurs at lower
values of $Y_e$.

It can be seen in addition that, as noticed previously by
\citet{Harris2018, Alford2021}, $\beta$-equilibrium in neutrino transparent
matter at non zero temperature does not occur at the point where
chemical potentials fulfil the usual $\beta$-equilibrium condition
for cold neutron stars ($\mu_n = \mu_p + \mu_e$ which stems from the
Fermi surface approximation, indicated by the red dashed line in
Fig.~\ref{fig:beta_eq_elastic}) but depends explicitly on the
neutrino production rates instead. Let us stress that, following
Fig.~\ref{fig:beta_eq_elastic}, only the rates including full
kinematics allow to recover the Fermi surface approximation at low
temperatures and thus the correct $\beta$-equilibrium conditions upon
evolving from a PNS to a neutron star.
\begin{table*}
  \begin{tabular}{|c||c|c|c||c|c|c|c|c|c|c|c|}
    \hline
  $t$&  $T$& $n_B$ & $Y_e$ &EoS & $x_n$ & $x_p$ & $U_n$ & $U_p$ & $\Delta U$& $m_n^*$ & $m_p^*$  \\
    (s)& (MeV) & (fm$^{-3}$)& & & & & (MeV) & (MeV) &(MeV) & (MeV)& (MeV) \\ \hline
    \hline
   0& 15.0& 0.250&0.30&RG(SLy4)&0.704 &0.296& 345.8 &258.7 &6.03 & 519.7 &599.5 \\
    &26.9& 0.017&0.22&RG(SLy4)& 0.702& 0.155&39.05 &7.586 &12.16 &893.7 &911.7 \\
    &11.2& 6.58 $\cdot 10^{-3}$&0.12&RG(SLy4)&0.751 &0.051&16.24 &-1.079 &7.30 &921.1 & 929.9\\
    &4.46& 9.35 $\cdot 10^{-6}$&0.34&RG(SLy4)&0.655 &0.339&5.772$\cdot 10^{-3}$ &-8.918$\cdot 10^{-3}$ &8.025$\cdot 10^{-3}$ & 939.5&938.3 \\ \hline
    0.7&20.1& 0.319&0.29&RG(SLy4)& 0.710&0.289&399.9 &321.9 & -8.237&462.2 &547.1 \\
    &35.7& 0.083&0.12&RG(SLy4)& 0.868&0.108& 188.7& 54.25&39.57 &723.5 & 817.2\\
    &18.8& 0.033&0.12&RG(SLy4)& 0.751 &0.055& 75.20&6.480 & 24.31 &854.3 & 897.4\\
    &4.18& 7.86 $\cdot 10^{-4}$&0.06&RG(SLy4)&0.848 &0.015&2.229 & -0.615&1.384 & 937.1& 937.3\\ \hline
    5.1& 48.6& 0.439&0.07&RG(SLy4)&0.931 &0.069& 475.9&375.4 &-80.93 & 353.5& 533.6\\
    &27.6& 0.256&0.07&RG(SLy4)& 0.931&0.069&379.3 & 193.8&11.48 & 477.8& 650.6\\
    &16.3& 0.150&0.07&RG(SLy4)&0.932 &0.068&284.1 & 97.58&39.46 &599.7 & 745.5\\
    &3.54& 8.58 $\cdot 10^{-3}$ &0.06&RG(SLy4)& 0.637& 5.01 $\cdot 10^{-5}$ &17.82 & -2.541& 8.494& 919.7&930.3 \\ \hline
    13.1&9.70& 0.514&0.06&RG(SLy4)& 0.939&0.061&487.4 & 443.8&-138.9 & 319.3& 500.5\\
    &5.71& 0.314&0.06&RG(SLy4)& 0.941&0.059&407.8 & 239.8&-16.09 & 429.2& 612.0\\
    &3.75& 0.196&0.05&RG(SLy4)& 0.946&0.054& 329.0& 130.8&29.96 & 537.6& 704.5\\
    &1.98& 0.016&0.05&RG(SLy4)& 0.657&1.422$\cdot 10^{-8}$& 33.62& -2.997& 14.20&901.8 &923.0 \\ \hline
\hline
  \end{tabular}
  \caption{Effective masses, interaction potentials and fractions of
    protons $x_p$ and neutrons $x_n$ under the thermodynamic
    conditions for which the opacities are shown in
    Figs.~\ref{fig:murcanu} and \ref{fig:murcanubar}. For each given time, the first line corresponds to the smallest radius and the last one to the largest radius. 
    $\Delta U = m^*_n - m^* _p
    + U_n - U_p - (m_n - m_p)$ is the shift in reaction threshold due
    to mean field effects, see \citet{Oertel2020}.
  }\label{tab:uint}
\end{table*}

\begin{figure*}
  \centering
  \includegraphics[width=\textwidth]{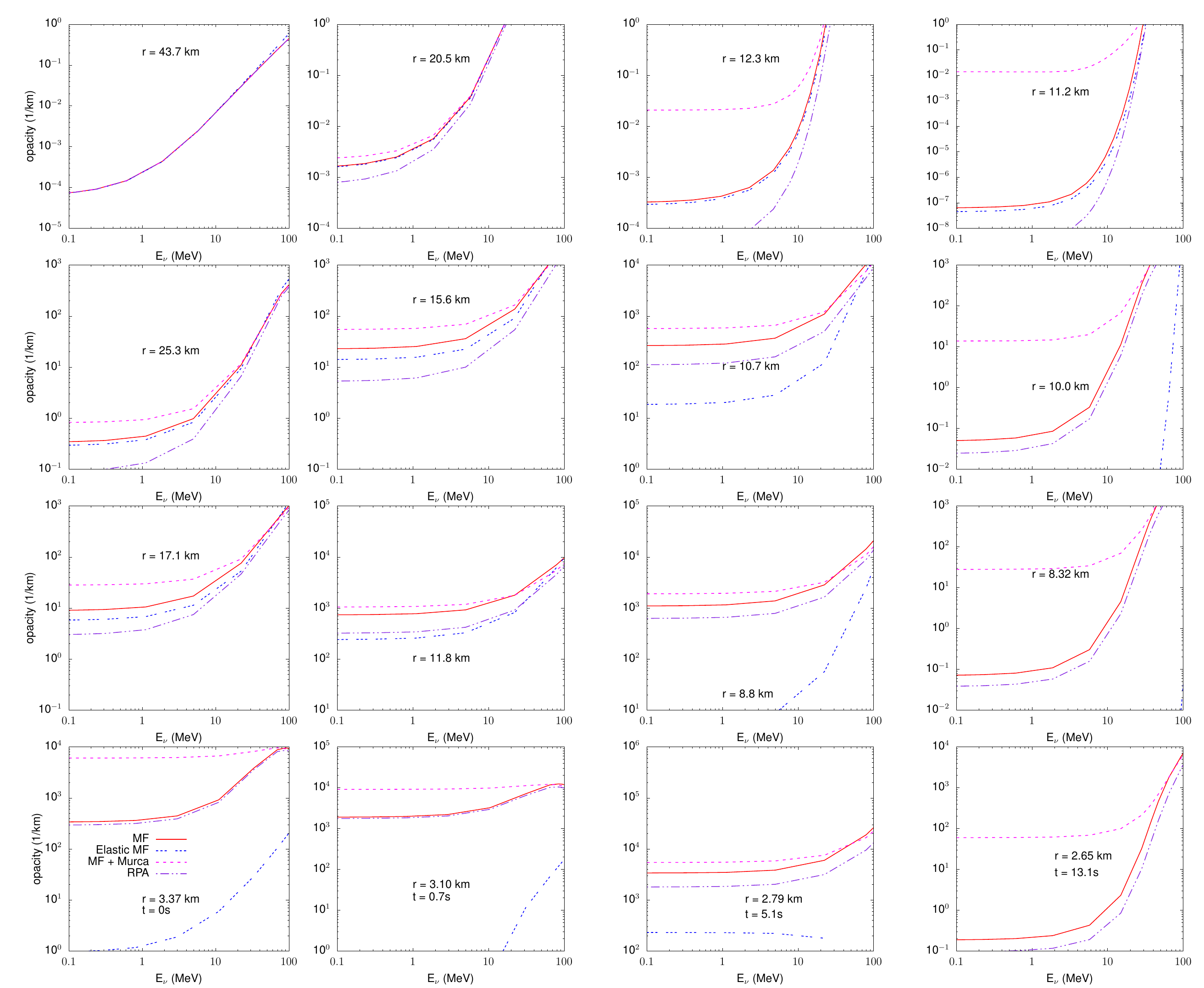}
  \caption{Neutrino ($\nu_e$) opacities with the RG(SLy4) EoS using
    different approximation schemes for calculating the reaction
    rates. The different thermodynamic conditions, see Table
    \ref{tab:uint} for the exact values, have been obtained from our
    fiducial simulation with the RG(SLy4) EoS and MF rates including
    MLT effects at different timesteps (increasing from left to right)
    and different locations (radius is increasing from bottom to top)
    in the star as indicated in the different panels. The uppermost
    panels correspond to conditions close to the surface of the star
    for that fiducial simulation.}
  \label{fig:murcanu}
\end{figure*}

\begin{figure*}
  \centering
  \includegraphics[width=\textwidth]{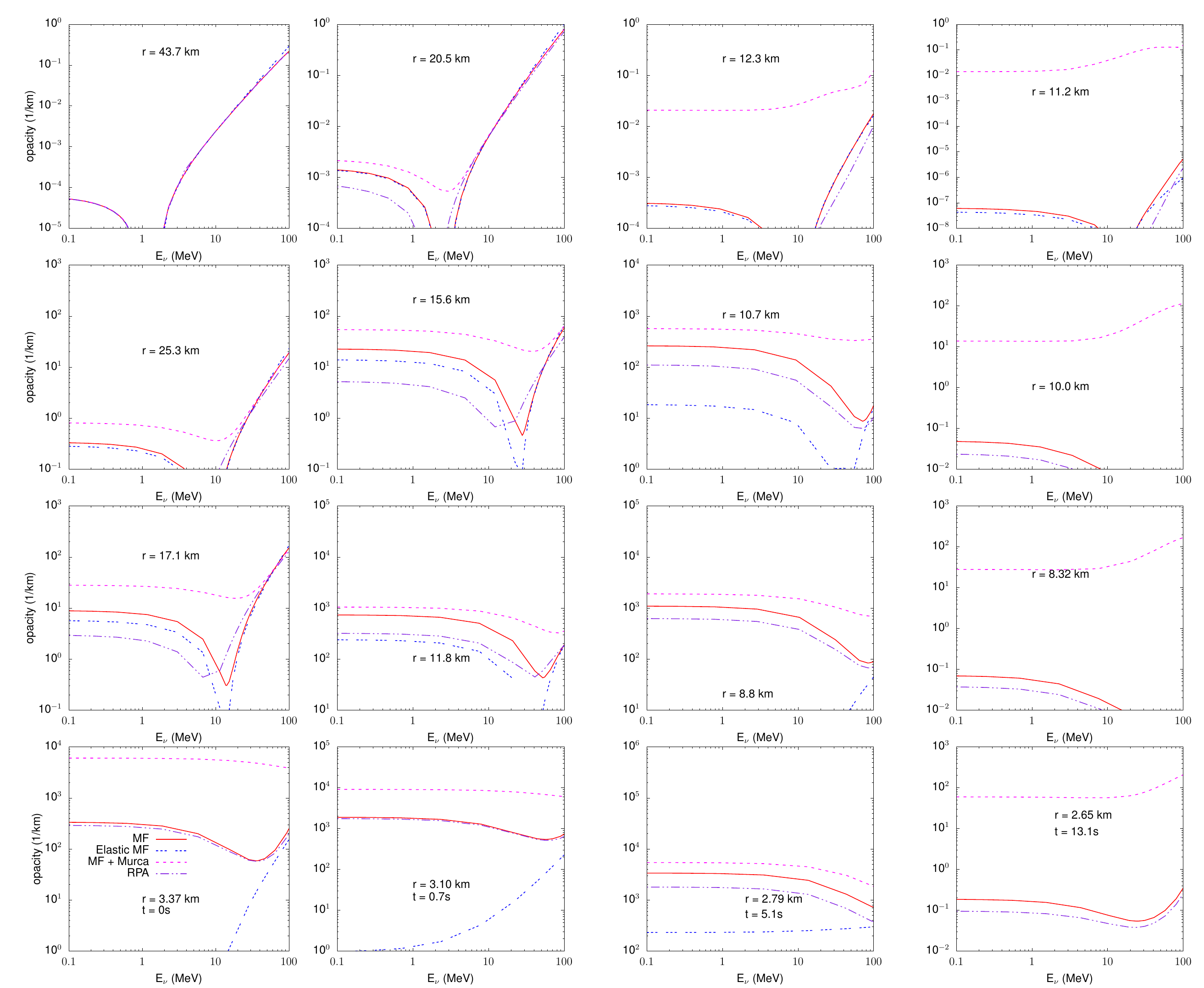}
  \caption{Same as Fig.~\ref{fig:murcanu} for antineutrino ($\bar{\nu_e}$) opacities. }
  \label{fig:murcanubar}
\end{figure*}

\subsection{Modified Urca processes}

From neutron star cooling, it is well known that under degenerate
conditions in cold neutron star matter the direct Urca processes, see
Eq.~\eqref{eq:durca}, are kinematically forbidden unless the proton
fraction exceeds roughly 10\%. In the same way, during the late stages
of PNS evolution they can become strongly suppressed for some neutrino
energies. In this case, the so-called modified Urca (mUrca) processes
become relevant, see also the discussion of $\beta$-equilibrium in
the hot merger remnant in \citet{Harris2018} and
\citet{Alford2021}. They involve a spectator nucleon $N$ allowing to
lift the kinematic restrictions of the direct processes :

\begin{align*}
  p + e^- + N \leftrightarrows n + \nu_e + N \qquad p  +
  N\leftrightarrows n + e^+ + \nu_e  + N \\
  n + e^+  + N\leftrightarrows p + \bar\nu_e  + N\qquad n + N
  \leftrightarrows p + e^- + \bar\nu_e  + N\numberthis{}
  \label{eq:murca}
\end{align*}

The relevant reactions rates have been studied extensively for
degenerate conditions in cold neutron stars following the seminal work
by \citet{FrimanMaxwell}, whereas much less effort has been dedicated
to hot matter. Due to their importance for CCSN neutrino spectra, the
corresponding neutral current reaction rates have, however, been
studied in more detail \citep[see e.g.][]{Hannestad1997}. Here we will
follow the phenomenological approach of \citet{Roberts2012neutrinos},
who have adapted the general framework for neutral current reactions in
\citet{Lykasov2008} to the above charged current reactions.

The idea is that the excitation of two-particle states, as required to
describe the reactions in Eq.~\eqref{eq:murca}, leads to a collisional
broadening which can be incorporated as finite quasi-particle lifetime
$\tau$ in the nuclear response function entering the rate calculation
\citep[see][for details]{Roberts2012neutrinos}. Because of vector
current conservation, the vector current contribution to the neutral
current rate vanishes in the limit of zero-momentum transfer, so that
it is generally assumed that the axial current contribution
dominates. In order to take mUrca reactions into account, we have
therefore implemented a finite width for the quasi-particles only in
the axial channel with numerical values for the lifetime taken from
\citet{Bacca2012}, see appendix~\ref{app:murca} for more details.

Let us stress that modelling mUrca processes in this way should not be
considered as quantitatively reliable. Among others, the momentum
transfer for the charged-current processes in dense matter is not
negligible, so that the contribution of the vector channel merits
further investigation. The values for the quasi-particle lifetime in
\citet{Bacca2012} have been obtained in the limit of vanishing momentum
transfer and for neutron matter for some selected temperatures and
baryon number densities. They can thus only be considered as a
guideline when applied to the entire temperature, electron fraction
and density domain needed for simulating PNS cooling. The main
physical effect should, however, be covered: as mentioned above, in
regions where direct Urca reactions are allowed, they give the
dominant contribution to (anti-)neutrino opacities and the collisional
broadening only marginally influences the opacities. The essential
effect of the mUrca processes as implemented here is thus to increase
opacities above/below thresholds for the direct processes.

As an example, we show opacities within the different approximation
schemes in Figs.~\ref{fig:murcanu} (neutrino) and \ref{fig:murcanubar}
(antineutrinos) employing the RG(SLy4) EoS. The thermodynamic
conditions for each panel correspond thereby to different times and
different radii inside the star as obtained from PNS profiles with the
fiducial simulation employing MF rates and the RG(SLy4) EoS including
MLT, see Sec.~\ref{sec:results}. The values for temperature, baryon
number density and electron fraction are listed in
Table~\ref{tab:uint}. In many cases, the antineutrino opacities
exhibit a pronounced threshold, where the increase in the opacity due
to the collisional broadening in this region is clearly visible.

\section{Simulation results}
\label{sec:results}

All simulations presented in this section are using the procedure
described in Section~\ref{sec:code}, and include --unless otherwise
stated-- convective effects modelled by MLT. We have first checked our
results with respect to the choice of the EoS, which has already been
discussed by several authors \citep[see e.g.][]{Keil1995,
  Sumiyoshi1995,Pons1999}. Among others, the influence of the symmetry
energy on the PNS cooling timescale has been largely
discussed~\citep{Sumiyoshi1995, Roberts2012convection,
  Nakazato:2019ojk}. We have performed simulations using the MF
prescription and three different EoS models,
RG(SLy4)~\citep{Gulminelli2015}, HS(DD2)~\citep{Hempel2010} and
SRO(APR)~\citep{Schneider2019}. All three of them have very similar
symmetry energies at saturation, $J = 32;31.7;32.6$ MeV, respectively,
and different slopes, $L = 46; 55; 58$ MeV.

We clearly confirm previous results in the literature and, in particular
by~\citet{Roberts2012convection}, according to which the time until
the convection stops and the luminosity drops decreases with the
symmetry energy slope of the equation of state. This indicates that
the faster contraction of the PNS and in particular the convective
effects outweigh the larger difference in neutrino and antineutrino
opacities. Let us stress here, too, that although for the usually
considered electron and proton capture reactions smaller symmetry
energies indeed lead to larger difference in $\nu_e$ and $\bar{\nu_e}$
opacities (see e.g. the comparison for RG(SLy4) and HS(DD2) in
\citet{Oertel2020}), this is no longer true for the neutron decay
reaction and its inverse, whose importance for the low energy
antineutrino opacity has been pointed out in \citet{Fischer2020}.

Finally, let us mention that in contrast to previous works we have
used the same EoS, both in the PNS evolution simulations and for
computing the initial model. As expected \citep[see for instance the
discussion in][]{Pons1999}, this has only little influence on the
results for the long term PNS evolution.

\begin{figure*}
  \centering
  \subfloat[Temperature $T$]{\includegraphics[width=0.33\textwidth]
    {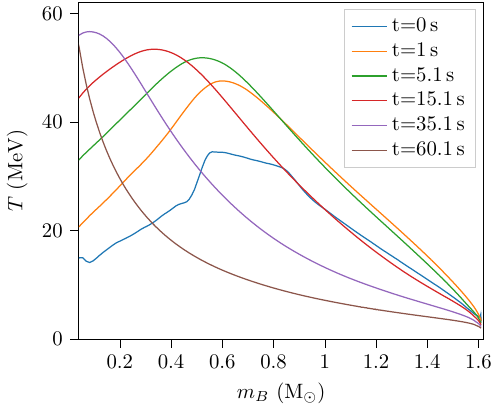}}
  \subfloat[Entropy per baryon
  $s$]{\includegraphics[width=0.33\textwidth] {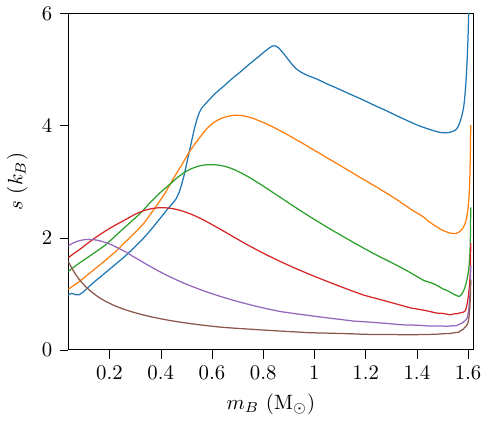}}
  \subfloat[Electron fraction $Y_e$]
  {\includegraphics[width=0.33\textwidth]{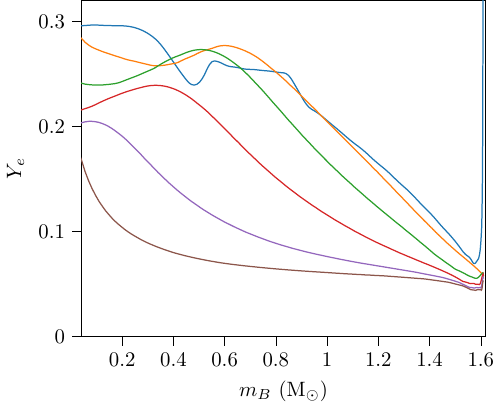}}
  \caption{PNS internal structure at selected times in our fiducial
    simulation without mixing length theory. Various relevant
    thermodynamic quantities are plotted as
    functions of the enclosed baryon mass $m_B(r) = m_N a(r)$.
  \label{fig:pns_ref_evol}}
\end{figure*}
\begin{figure*}
  \centering
  \subfloat[Temperature $T$]{\includegraphics[width=0.33\textwidth]
    {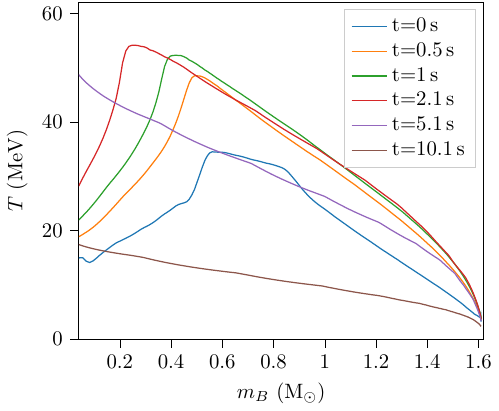}}
  \subfloat[Entropy per baryon
  $s$]{\includegraphics[width=0.33\textwidth] {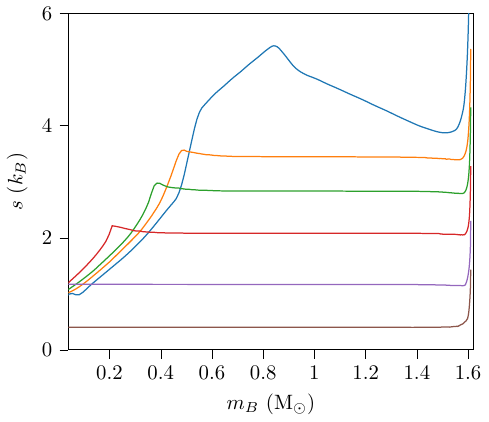}}
  \subfloat[Electron fraction
  $Y_e$]{\includegraphics[width=0.33\textwidth] {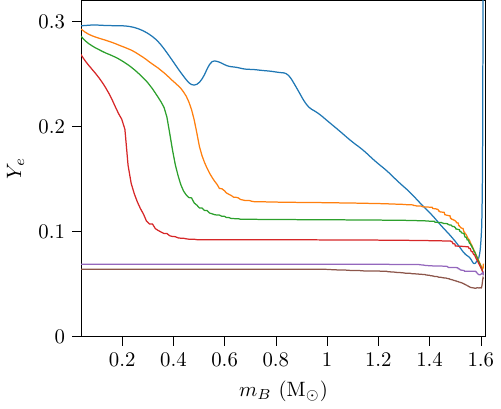}}
  \caption{Same as Fig.~\ref{fig:pns_ref_evol} including convective effects via the mixing length theory.
  \label{fig:pns_mlt_evol}}
\end{figure*}

\subsection{Influence of convection on proto-neutron star evolution}
\label{ssec:convec}
In order to show the importance of convective effects during PNS
evolution, we compare two simulations, with and without MLT. The EoS
used for both the CCSN evolution and the quasi-static PNS modelling is
RG(SLy4) \citep{Gulminelli2015} and the MF prescription for
charged-current neutrino reaction rates has been employed, as
described in Section~\ref{sec:cc_int} and in \citet{Oertel2020}.

Fig.~\ref{fig:pns_ref_evol} shows the radial profiles of temperature
$T$, entropy per baryon $s$ and electron fraction $Y_e$, as functions
of the enclosed baryon mass $m_B$ at different times in our fiducial
simulation which does not include convective effects.
Fig.~\ref{fig:pns_mlt_evol} shows the same profiles, but this time
including convective effects via the MLT. The profiles at
$t=\SI{0}{\second}$ correspond to our initial data, which are taken
at about \SI{500}{\milli\second} after bounce.

\begin{figure}
  \centering
  \includegraphics[width=0.85\columnwidth]{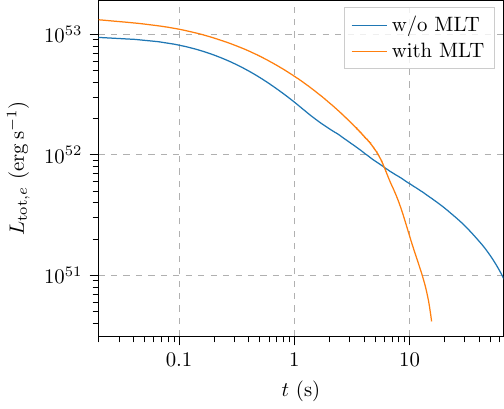}
  \caption{Total (sum of 6 flavors) neutrino energy luminosity as a
    function of time, in our fiducial simulation, with and without
    MLT.}
  \label{fig:pns_mlt_elum}
\end{figure}

It is obvious that convection in the mantle of the PNS is extremely
efficient: even by taking the same initial model as in our fiducial
simulation (obtained with a spherically symmetric hydrodynamic
simulation, thus without any convective effects) it takes less than
about \SI{100}{\milli\second} to obtain a uniform profile of entropy
per baryon $s$ and electron fraction $Y_e$ in the convectively unstable
zone, which is then maintained in a state close to neutral
buoyancy. These observations are qualitatively similar to what is
obtained in full hydrodynamic studies, as for example in the Figs.~7
and 18 of \citet{Nagakura2020}.

Another striking effect of including convective motions in the
simulations is the difference in the evolution timescales: the
non-convective model takes about four times longer than the convective
one to fully deleptonize. Indeed, convective motions in the PNS are
carrying leptons and heat from the inner boundary of the convective
layer to its outer part, much closer to the neutrinospheres. As a
consequence of the higher temperatures in the outer layers, the
neutrino luminosity is about \num{1.5} times higher with MLT than
without, as shown in Fig.~\ref{fig:pns_mlt_elum}. A sudden drop in the
luminosity in observed after about \SI{4}-\SI{5}{\second}, when the
star becomes homogeneous in entropy and electron fraction and the
convective motions stop.

\begin{figure}
  \centering
  \includegraphics[width=0.85\columnwidth]{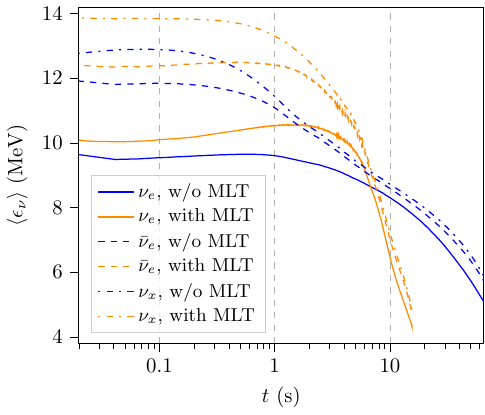}
  \caption{Mean energy of emitted neutrinos as a function of time, for
    each neutrino flavor, in our fiducial simulation, with and without
    MLT.}
  \label{fig:pns_mlt_emean}
\end{figure}

The emitted neutrino spectrum is also significantly modified by
convection. Fig.~\ref{fig:pns_mlt_emean} represents the mean energy
of emitted neutrinos (as measured by a distant observer). We see that
the mean energy of all neutrino flavors is globally enhanced during
the early evolution with MLT, and then quickly drops when the star reaches
neutral buoyancy and convection stops.

It should be stressed that we do not recover the results of Fig.~20 of
\citet{Nagakura2020}, where the authors observed that at early times
the convective effects decrease the mean energy of emitted
neutrinos. We do not believe our simulation to be accurate enough at
that early epoch, as we neglected a lot of important physical
ingredients such as the accretion onto the PNS, which is known to
significantly alter the neutrino signal in particular by causing a
time variation \citep[see][]{Nagakura:2021lma}. Instead, we focus on the
long term evolution after the shock departure, for which our
observations seem to be robust and confirm previous findings on the
flattening of $s$ and $Y_e$ profiles as well as the luminosity, see
e.g.~\citet{Roberts2012convection} and \citet{Roberts2017handbook}.

In addition to the potentially observable drop in neutrino
luminosities \citep{Roberts2012convection}, the modifications in the
neutrino spectra due to convective effects are quite important in the
study of the neutrino driven wind (NDW). We estimate the electron
fraction using the approximation developed in \citet{Qian1996},
\begin{equation}
  Y_e^{\text{NDW}} = \left[ 1 + \frac{L_{\bar\nu_e,e}
      (\varepsilon_{\bar\nu_e} -2\Delta +
      \num{1.2}\Delta^2/\varepsilon_{\bar\nu_e}) }{L_{\nu_e,e}
      (\varepsilon_{\nu_e} + 2\Delta +
      \num{1.2}\Delta^2/\varepsilon_{\nu_e}) }
  \right]^{-1} \label{eq:ye_ndw}
\end{equation}
where $L_{\nu_e,e}$ and $L_{\bar\nu_e,e}$ are the electron neutrino
and antineutrino energy luminosities, $\Delta = m_n c^2 - m_p c^2$ is
the neutron-proton mass difference, and $\varepsilon_\nu$ is defined
as
$\varepsilon_\nu = \langle \epsilon_\nu^2 \rangle / \langle
\epsilon_\nu \rangle$ where $\epsilon_\nu$ is the energy of a given
neutrino flavor $\nu$. This approximation considers only the processes
$n + \nu_e \leftrightarrows p + e^-$ and
$p + \bar\nu_e \leftrightarrows n + e^+$, in the elastic approximation
neglecting the mass of the electron $m_e$ and the Pauli blocking
effect of leptons on the final state.

It should be stressed that the formula (\ref{eq:ye_ndw}) is obtained
by considering that the NDW is composed of free neutrons and protons
only. By doing so the so-called \textit{alpha effect} is neglected,
which results from the formation of $\alpha$ particles and can
introduce significant changes in the electron fraction \citep[see
e.g.][]{Meyer1998}. But as our goal is only to estimate the global
effect of modified neutrino spectra due to convective effects on the
NDW, the formula (\ref{eq:ye_ndw}) should nevertheless give us the
global trend.

\begin{figure}
  \centering
  \includegraphics[width=0.9\columnwidth]{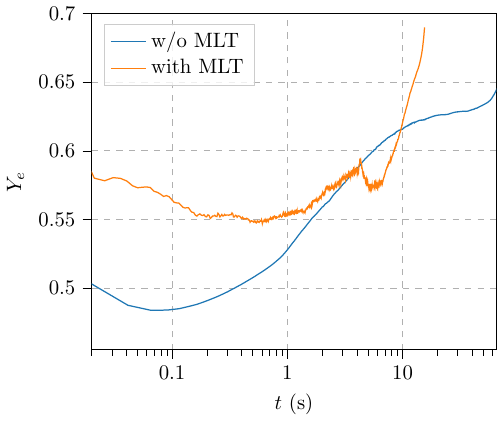}
  \caption{Electron fraction in the neutrino driven wind, computed
    using Eq.~(\ref{eq:ye_ndw}), in our fiducial simulation, with and
    without MLT.}
  \label{fig:pns_mlt_yendw}
\end{figure}

The evolution of the electron fraction in the NDW is represented in
Fig.~\ref{fig:pns_mlt_yendw}. We see that in our simulations
convective effects push the composition of the NDW to higher electron
fractions and that it is always proton-rich. The structure which can
be observed in the simulation with MLT at about \SI{4}-\SI{5}{\second}
results from convection ceasing.  Most recent CCSN simulations
indicate a proton-rich NDW and under such conditions the most probable
nucleosynthesis processes occurring in the wind are the weak r-process
and the $\nu p$-process \citep[see e.g.][]{Arcones2013}.

\subsection{Influence of neutrino interaction rates on the PNS evolution}
\label{ssec:cc}

We study the influence of the various prescriptions for charged
currents presented in Section~\ref{sec:cc_int} by performing
simulations using either
\begin{itemize}[label=\textbullet]
  \item the elastic approximation with Mean Field corrections, without mUrca effects, denoted Elastic MF
  \item the Mean Field prescription without mUrca effects, denoted MF
  \item the Mean Field prescription with mUrca effects, denoted MF+mUrca
  \item the RPA $t_3'$ prescription without mUrca effects, denoted RPA $t_3'$~,
\end{itemize}
employing again the RG(SLy4) EoS \citep{Gulminelli2015}. 

\begin{figure*}
  \centering
  \subfloat{\includegraphics[width=0.45\textwidth]{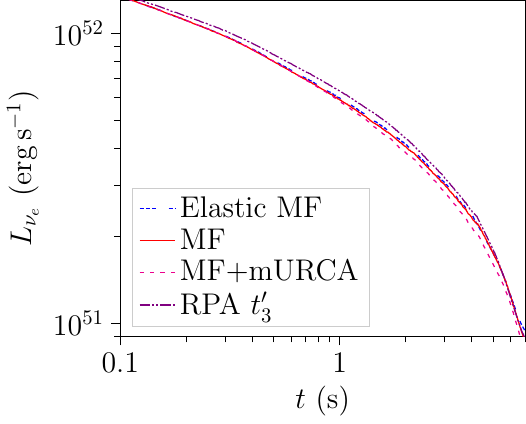}}
  \subfloat{\includegraphics[width=0.47\textwidth]{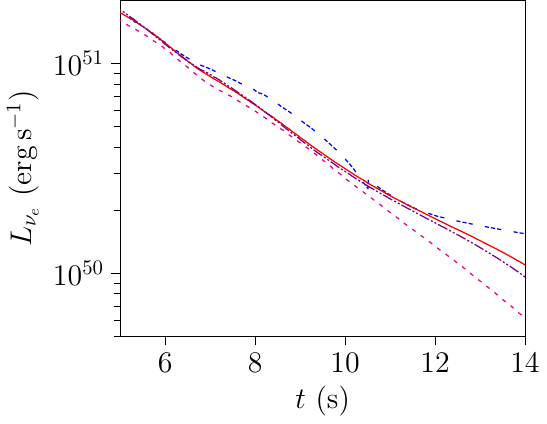}}
  \caption{Energy luminosity of emitted electron neutrinos as function of time, for the four different prescriptions to compute charged current neutrino-nucleon interactions}
  \label{fig:pns_rpa_lum}
\end{figure*}

We see little impact from these different prescriptions on the
early PNS evolution during the convective phase. Some
differences start to appear when convection stops and the luminosity
starts to drop, as shown in Fig.~\ref{fig:pns_rpa_lum}, which
shows the evolution of electron (anti-)neutrino luminosities as
functions of time.

The differences caused by the elastic approximation start to be
significant at this point, and the electron neutrino luminosity is
much higher. This is explained by two factors: since the rates are
lower the transport of energy and leptons in the PNS is more
efficient, and the displacement of $\beta$-equilibrium towards more
neutron rich matter leads to a more important deleptonization.

Regarding the effect of collisional broadening, the neutrino
luminosities in simulations including mUrca are always lower than in
those which do not. This can easily be understood. Since the opacities
are globally enhanced by collisional broadening, the diffusion of heat
and leptons inside the PNS is less efficient and the
neutrinosphere is located at a lower temperature. This difference
becomes more and more pronounced after about \SI{10}{\second} of
evolution, when the temperature starts to drop, enhancing
the differences between the prescriptions.

\begin{figure}
  \centering
  \subfloat[Electron neutrino $\nu_e$]{\includegraphics[width=0.8\columnwidth]{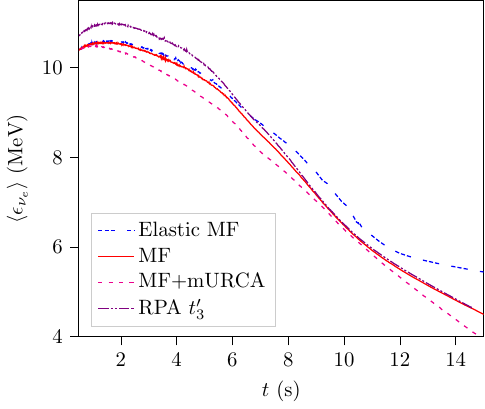}}\\
  \subfloat[Electron antineutrino $\bar\nu_e$]{\includegraphics[width=0.8\columnwidth]{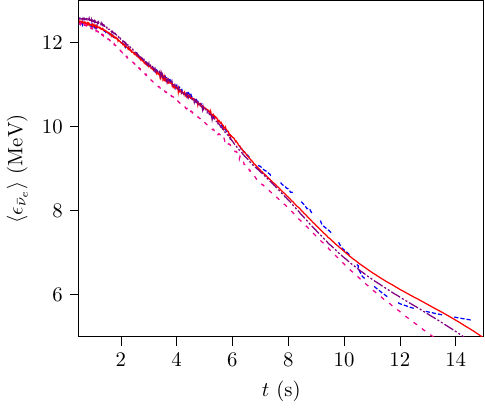}}
  \caption{Mean energy of emitted electron (anti-)neutrinos as
    functions of time for the four different prescriptions for charged
    currents.}
  \label{fig:pns_rpa_emean}
\end{figure}

We can note here that, when comparing MF and MF with mUrca
simulations, there are little differences in the energy luminosity
(Fig.~\ref{fig:pns_rpa_lum}) and in the mean energy of emitted
neutrinos (Fig.~\ref{fig:pns_rpa_emean}), whereas opacities are quite
different, as seen from Figs.~\ref{fig:murcanu} and
\ref{fig:murcanubar}. This can be explained from the fact that the
large differences in the rates occur in regions which are finally not
so relevant for the neutrino emission and the overall evolution is
dominated by other effects. In particular, the cooling of the central
region is dominated by convective mixing whose dynamics are
essentially independent of the charged currents. The exact location of
the neutrinospheres is also not only determined by charged-current
opacities but probably dominated by the scattering opacity which does
not change between the different simulations. In addition, close to
the neutrinospheres, relevant for neutrino emission, the differences
in charged-current opacities are much smaller than in the centre or
well above the neutrinospheres. We have checked this in our
simulations, comparing the opacities for relevant thermodynamic
conditions obtained from our fiducial run with MF rates in different
regions of the PNS, see Figs~\ref{fig:murcanu} and
\ref{fig:murcanubar}. Close to the neutrinospheres, the difference in
opacities with or without taking into account mUrca are not so large.

Finally, simulations which include RPA effects have a slightly higher
luminosity during the first phase of the evolution. This can be
understood because the reduced opacities due to nuclear correlations
improves the diffusion of heat in the PNS. At later times this trend
is inverted, since the model with RPA is colder and the neutrino mean
free path becomes of the order of the size of the star. This behaviour
is similar to the observations made in Fig.~5 of
\citet{Roberts2017handbook}, but some important differences can be
noticed. Among others, in \citet{Roberts2012convection,
  Roberts2017handbook} differences between RPA and MF start to appear
only after convection has stopped, and become significant at late
times, whereas our simulations show only small differences at late
times. These discrepancies might have several origins. Differences
coming from the initial model -- we recall that we compute the initial
model consistently with the same EoS and reaction rates as the
simulation -- could explain these different behaviours, as our RPA and
MF results might differ at early times in contrast to those in
\citet{Roberts2017handbook}. As discussed in \citet{Oertel2020}, the
importance of RPA correlations is energy dependent and for some
(anti)-neutrino energies almost no difference is observed with the MF
opacities. Therefore, the inclusion of this energy dependence in our
work in contrast to \citet{Roberts2012convection,Roberts2017handbook}
is another possible explanation for the discrepancies. We expect the
average difference between RPA and MF to be reduced for the full rates
compared with a grey factor, which seems indeed to be the case. We
also include the full inelasticity of neutrino-nucleon scattering,
which leads to faster equilibration of neutrino spectra and thus
reduces the differences between the different prescriptions for the
charged current rates. The treatment of neutrino transport or the EoS
certainly play a role, too.


Let us now have a look at the emitted neutrino
spectrum. Fig.~\ref{fig:pns_rpa_emean} represents the mean energy of
emitted electron (anti-)neutrinos (as measured by a distant observer).
The findings are similar to those discussed above, the differences
introduced by the elastic approximation are the most significant at
late times, especially for electron neutrinos with enhanced mean
energies. mUrca processes systematically lower the mean energy with
increasing differences again appearing around \SI{10}{\second} of
evolution. Globally, as for the luminosities, the inclusion of mUrca
processes has a stronger impact on the mean energies than nuclear
correlations included via RPA. The latter produces a stronger
enhancement of neutrino mean energy and, to a lower extent, that of
anti-neutrinos at early times. Here again, the trend is inverted at
later times with mean energies of both electron neutrinos and
antineutrinos slightly smaller with the RPA approach than in MF.

\begin{figure}
  \centering
  \includegraphics[width=0.9\columnwidth]{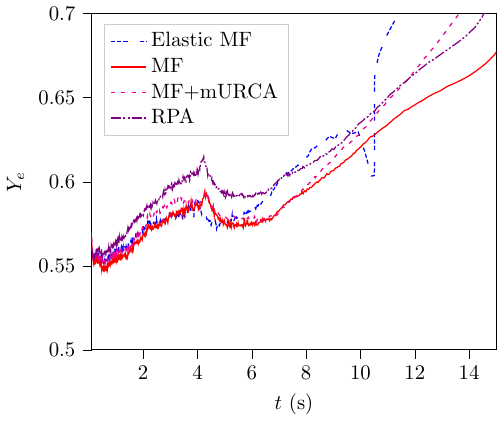}
  \caption{Electron fraction in the neutrino driven wind, computed
    with \eqref{eq:ye_ndw}, as a function of time, for the four
    different prescriptions for charged currents.}
  \label{fig:pns_rpa_yendw}
\end{figure}

As far as the impact on the NDW is concerned,
Fig.~\ref{fig:pns_rpa_yendw} (left panel) shows the evolution of the
electron fraction in the NDW comparing the different prescriptions for
computing charged current rates.
\begin{figure*}
  \centering
  \includegraphics[width=0.9\columnwidth]{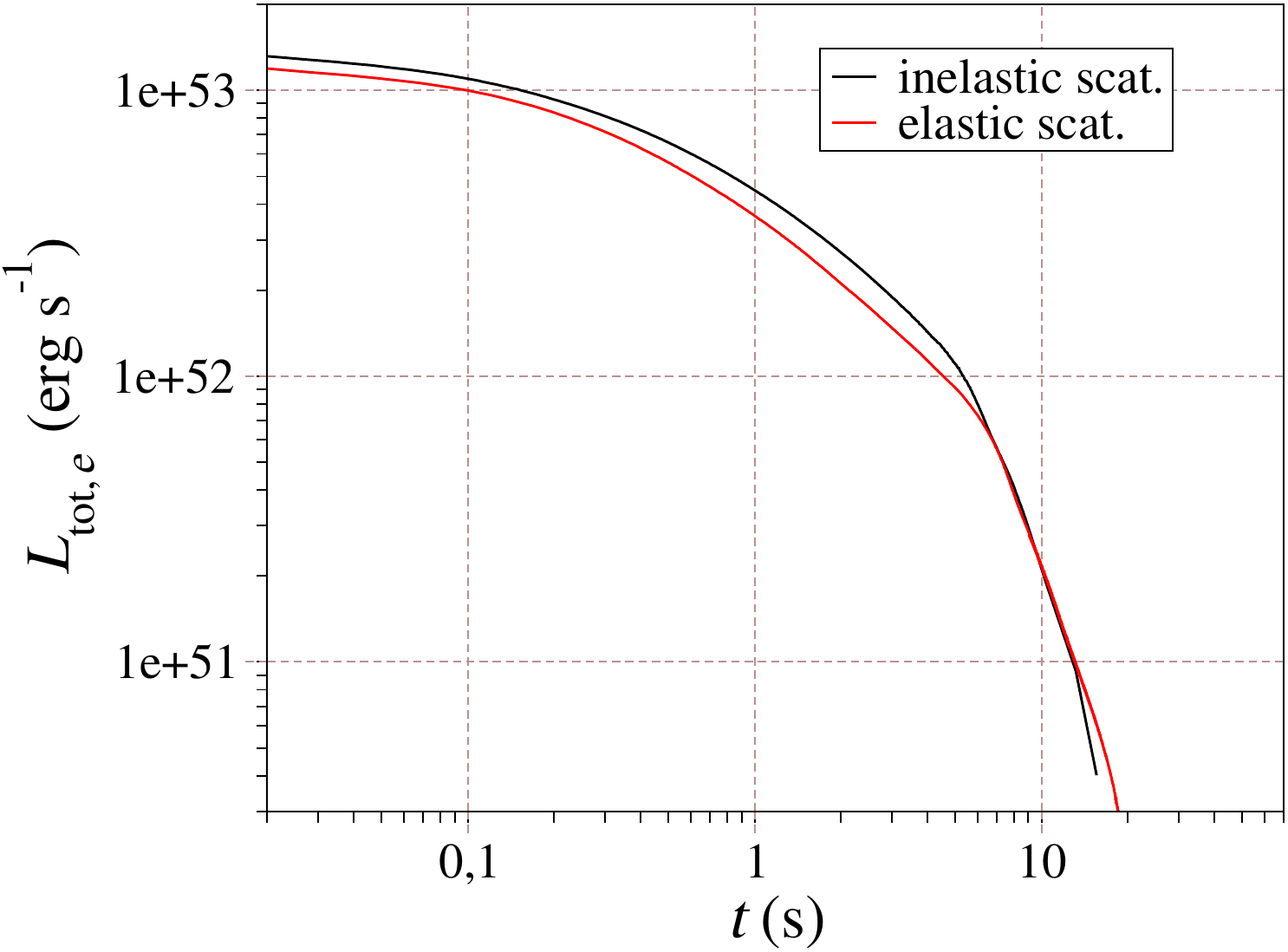}\hfill
  \includegraphics[width=0.9\columnwidth]{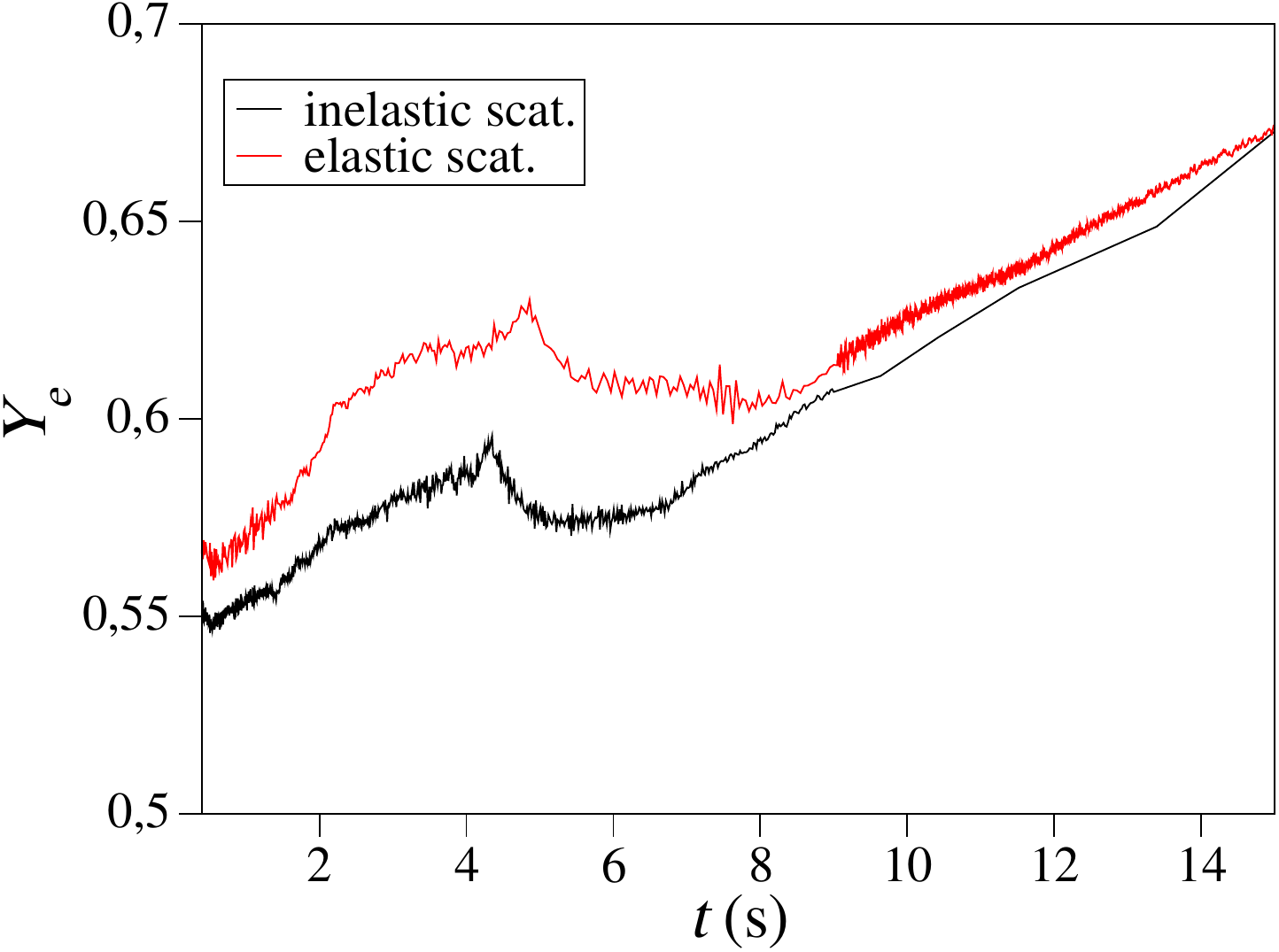}
  \caption{Total (6 flavors) neutrino energy luminosity (left) and
    electron fraction in the neutrino driven wind (right) computed
    with \eqref{eq:ye_ndw}, as functions of time, comparing the
    elastic approximation \citep{Bruenn1985} for $\nu$-$N$ scattering
    with the inelastic rates from \citep{Thompson2000}. The RG(SLy4)
    EoS and MF charged current rates have been employed.}
  \label{fig:pns_scat}
\end{figure*}
It can be seen that the stronger enhancement within RPA at early times
of the electron neutrino average energy, with respect to
antineutrinos, (see Fig.~\ref{fig:pns_rpa_emean}) has some noticeable
consequences on the electron fraction in the NDW, which is about
\SI{4}{\percent} higher at early times than in MF. mUrca processes
which essentially affect neutrinos and antineutrinos in the same way,
show only little influence on the composition of the NDW at early
times. The prediction of simulations including mUrca start to be
different after about \SI{8}{\second}, increasing the electron
fraction. Note, however, that when determining $Y_e$ in the NDW with
Eq.~(\ref{eq:ye_ndw}), we have used several approximations. In
particular we have assumed only free nucleons, which should be
considered with caution over such a long timescale.

Although in this paper we focus on the impact on charged current
rates, it should be noted that in most parts of the PNS,
neutrino-nucleon scattering is the dominant source of opacity and
different ways of computing the corresponding rates are expected as
for the charged current rates to change the opacities and thus the PNS
evolution. We will not perform here a detailed study, but as
indication of the impact different scattering rates can have, we
compare a simulation employing the elastic rates from
\cite{Bruenn1985} with our fiducial simulation using fully inelastic
rates from \cite{Thompson2000}, see
appendix~\ref{appendix:col_int}. The overall evolution is again rather
similar, see Fig.~\ref{fig:pns_scat} where as an example we show the
total neutrino energy luminosity and the electron fraction in the
neutrino driven wind. At early times the elastic rates predict
a slightly lower neutrino luminosity and a slightly more proton rich matter
matter in the neutrino driven wind, but the evolution is again
dominated by convection effects.

\section{Summary and discussion}
\label{sec:summary}

In this study we have performed simulations of the Kelvin-Helmholtz
cooling phase of PNSs with a new code relying on the
quasi-static approximation. Our code is modelling neutrino
transport with the Fast Multigroup Transport (FMT) radiation scheme
and includes convective effects within the mixing length theory (MLT).

We motivated the use of the mixing length theory by performing
simulations with and without MLT, and conclude that simulations with
convection yield results qualitatively different from simulations
without. The PNS contracts much faster because of an efficient heat
and lepton transport from inner regions to the neutrinospheres, the
energy of emitted neutrinos is higher and the NDW is more
proton-rich. As already discussed earlier, when convection stops there
is a clear break in the neutrino luminosity, potentially observable.

In the continuation of our previous work on the influence of
charged-current neutrino-nucleon interactions in core-collapse
simulations \citep{Oertel2020} we studied the influence of nuclear
correlations and collisional broadening due to modified Urca processes
in the evolution of PNSs. We recall that commonly used
approximations in CCSN (the elastic approximation) and in NS cooling
simulations (the Fermi surface approximation) are not compatible and
that PNS evolution simulations should rely on charged current rates
computed with full kinematics.

From our simulations we conclude that the addition of mUrca has a
certain, though limited effect on the neutrino emission, by reducing
both the mean energy and the luminosity of emitted electron
(anti-)neutrinos. Moreover, the mUrca processes become dominant at low
temperatures. This can be seen from Figs.~\ref{fig:murcanu} and
\ref{fig:murcanubar}, where we have taken the thermodynamic conditions
from PNS profiles obtained with the fiducial simulation employing MF
rates and the RG(SLy4) EoS at different times and at different radii
inside the star.  The highest radius at each given time is located
close to the surface.  It is obvious from that figure that at later
times and closer to the surface, opacities from mUrca reactions
dominate. A similar effect is observed for neutrinos. Thus although
neutrino transparency has not been reached (the neutrino mean free
path is still smaller than the size of the star) at the end of the
simulation, we can confirm that mUrca processes dominate the late
evolution when direct processes become kinematically strongly
suppressed.

The inclusion of nuclear correlations via RPA mostly has an effect at
the beginning of the simulation, by enhancing the mean energy of
emitted electron neutrinos, which indirectly increase the electron
fraction in the NDW. At later times our simulations with RPA show
little difference in the results, with respect to mean field
rates. Thus although qualitatively in agreement with previous works
\citep[see e.g.][]{Roberts2017handbook}, there are some important
quantitative discrepancies. As discussed in Sec.~\ref{ssec:cc}, there
are several possible reasons for that: differences in the initial
model, the energy dependence of the RPA rates included in our work,
the faster equilibration of neutrino spectra due to inelastic
scattering off nucleons, the neutrino treatment or the EoS. These
different effects should be investigated further before any generic
conclusion on the impact of nuclear correlations on PNS evolution and
the resulting neutrino emission can be drawn.

Of course our study has several limitations. The simulations
investigating the different prescriptions for charged current
neutrino-nucleon interactions have been performed with only one EoS,
thus we cannot exclude that other EoS show different effects. In
addition, although we do not see much difference in the results from
different progenitors, it should be mentioned that only a few have
been used and that some particular progenitors might show different
behaviour. The numerical method used could be improved, too. The
neutrino transport scheme is an approximation to the full Boltzmann
transport. It has the advantage of being computationally not very
expensive and at the same time being more elaborate than standard
equilibrium flux limited diffusion methods. In particular in the
semi-transparent regime it behaves as well as some recently employed
more sophisticated, but computationally more expensive methods \citep[as
e.g. variable Eddington transport methods in][]{Roberts2012code}. The
transition between the CCSN evolution code with full-hydrodynamics and
our quasi-static PNS evolution code induces a small discontinuity in
some quantities, which, however, should only impact the very early
evolution. As again shown here, convection plays a very important role
for PNS evolution. The MLT scheme should govern the main qualitative
features, but only a more detailed multi-dimensional study can fully
account for convective effects. In addition, we have completely
neglected the accretion process, therefore the first few hundreds of
milliseconds in the evolution of our models should not be considered
as reproducing accurately the conditions in CCSN.

We nevertheless demonstrated the feasibility of PNS studies including
mixing length theory and state-of-the-art microphysics for the
computation of neutrino interactions with a new computationally
low-cost numerical algorithm. This kind of method allows to study a
wide parameter space in a reasonable time and will certainly prove
useful as uncertainties in PNSs evolution are still numerous.

\section*{Acknowledgements}

We would like to thank B. Müller for providing us with the original
FMT code and Tobias Fischer for discussions. The research leading to
these results has received funding from the PICS07889 and the
Observatoire de Paris through the action fédératrice ``PhyFog''. This
work was granted access to the computing resources of MesoPSL financed
by the Region Ile de France and the project Equip@Meso (reference
ANR-10-EQPX-29-01) of the programme Investissements d’Avenir
supervised by the Agence Nationale pour la Recherche. The authors
gratefully acknowledge the Italian Istituto Nazionale de Fisica
Nucleare (INFN), the French Centre National de la Recherche
Scientifique (CNRS) and the Netherlands Organization for Scientific
Research for the construction and operation of the Virgo detector and
the creation and support of the EGO consortium.

\section*{Data Availability}

No new data were generated or analysed in support of this research.



\bibliographystyle{mnras}
\bibliography{biblio} 


\begin{appendix}

\section{Fast Multigroup Transport}
\label{appendix:fmt}

In a spherically-symmetric spacetime described by the metric
\eqref{eq:metric}, the stationary Boltzmann equation can be written as
\begin{align*}
  \frac{\mu}{\psi} \frac{\partial f}{\partial r} + \frac{\mu \, \epsilon}{\psi}(\partial_r \ln\alpha)  \frac{\partial f}{\partial \epsilon} \\+ \left( \frac{1}{r} - (\partial_r \ln\alpha) \right) \frac{1-\mu^2}{\psi} \frac{\partial f}{\partial \mu} = \mathcal{B}[f]
\end{align*}
where $f(r,\epsilon,\mu)$ is the distribution function, $\epsilon =
p^0$ is the energy of neutrinos and $\mu = \cos\Theta = p^r / p^0$ is
the cosine of the propagation angle.  In the stationary case we can
treat the redshift by a simple change of variable : we introduce the
redshifted energy $\hat \epsilon = \alpha \epsilon$. For the function
$f(r,\hat \epsilon,\mu)$ the Boltzmann equation now simply becomes
\begin{equation}
  \frac{\mu}{\psi} \frac{\partial f}{\partial r} + \left( \frac{1}{r} - (\partial_r \ln\alpha) \right) \frac{1-\mu^2}{\psi} \frac{\partial f}{\partial \mu} = \mathcal{B}[f]
\end{equation}

Let us now introduce the three first angular moments of the
distribution function : $J = \frac{1}{2} \int f d\mu$,
$H = \frac{1}{2} \int f \mu d\mu$ and
$K = \frac{1}{2} \int f \mu^2 d\mu$ and write the collision integral
as $\mathcal{B}[f] = j - \chi f$, which is possible if we consider
only isotropic processes (see appendix \ref{appendix:col_int}). The
first two moment equations are then
\begin{align}
  \frac{\alpha^2}{r^2 \psi} \frac{\partial}{\partial r} \left( \frac{r^2}{\alpha^2} H \right) &= j - \chi J \label{eq:nu_mom_1} \\
   \frac{1}{\psi} \frac{\partial K}{\partial r} + \left( \frac{1}{r} - (\partial_r \ln\alpha) \right) \frac{3K - J}{\psi} &= - \chi H \label{eq:nu_mom_2}
\end{align}

The solution of this system is computed by using the fast neutrinos
transport scheme by \citet{Muller2015}. In the high optical depth area the solution is obtained using a two-stream approximation,
\begin{align*}
  \frac{1}{\psi} \frac{\partial f_{out}}{\partial r} &= j - \chi f_{out}
  -\frac{1}{\psi} \frac{\partial f_{in}}{\partial r} &= j - \chi f_{in}~.
\end{align*}
It should be stressed that these two equations are usually coupled
because the coefficients $j$ and $\chi$ depend of the neutrino
distribution function, though in our case we use the neutrino distribution of
the previous timestep to compute them (see appendix
\ref{appendix:col_int}). The flux factor $h=H/J$ is obtained by
assuming a continuous distribution $f(\epsilon,\mu) \propto
\mathrm{e}^{a\mu}$
\begin{equation}
  h = 1 +  \frac{2\,f_{in}/f_{out}}{1-f_{in}/f_{out}} + \frac{2}{\ln(f_{in}/f_{out})}
\end{equation}
and we can then solve the flux-divergence equation
\ref{eq:nu_mom_1}.

The solution at low optical depth is obtained with a
two-moment closure:
\begin{equation}
   k(h) = \frac{K}{J} = \frac{1 - 2h + 4 h^2}{3}
\end{equation}
which can be used to transform the equation \ref{eq:nu_mom_2} into an
ordinary differential equation for the flux factor
\begin{align*}
  \frac{dh}{dr} = \frac{1}{k - h k'(h)} \left\lbrace \left(
  \frac{1}{r} - \partial_r\ln\alpha \right)(k-1)h - \right.\\
  \left. \psi \left( \chi h^2 -k  \frac{j}{J} + k \chi \right)
  \right\rbrace \numberthis{}~.
\end{align*}
These two solutions are matched at the point $h=\num{0.51}$, to avoid
the singular point $h=\num{0.5}$\,. 

\medbreak
Finally, the neutrino fluxes used in equations~\eqref{eq:nu_src_1} and
\eqref{eq:nu_src_2} are computed as 
\begin{align}
  F_{\nu,n}(r) &= \frac{4\pi c}{(hc)^3} \int H(r,\hat\epsilon) \frac{\hat \epsilon^2 d\hat\epsilon}{\alpha^3} \\
  F_{\nu,e}(r) &= \frac{4\pi c}{(hc)^3} \int H(r,\hat\epsilon) \frac{\hat \epsilon^3 d\hat\epsilon}{\alpha^4}~.
\end{align}

\section{Neutrino collision integral}
\label{appendix:col_int}

We limit ourselves to isotropic scattering and pair production
kernels, which allows us to recast the collision integral to the form
\begin{equation}
  \mathcal{B}[f] = j_{\mathrm{eff}} - \chi_{\mathrm{eff}} f~,
\end{equation}
where the coefficients $j_{\mathrm{eff}}$ and $\chi_{\mathrm{eff}}$
depend on the distribution function $f$. These effective coefficients
are computed using the value of $f$ from the previous timestep. This
approximation allows us to use the FMT algorithm
while having very little influence on the result. The fact that
interaction kernels are non-isotropic is usually taken into account by
including their first Legendre moments, but this is impossible within
the FMT. These anisotropies are expected to have only an influence in
the semi-transparent regime, and should not change our results
qualitatively.

\subsection{Scattering integral}

The dominant processes for scattering of neutrinos in dense nuclear
matter are scattering off free nucleons $\nu + N \leftrightarrows \nu
+ N$, and on a less important level the scattering off free
electrons/positrons $\nu + e^\pm \leftrightarrows \nu + e^\pm$. The
scattering off nucleons is treated as if the nucleons were an ideal
gas \citep[see e.g.][]{Thompson2000} with full inelasticity, and
electrons/positrons are treated as a relativistic ideal gas
\citep[see][]{Yueh1977, Chernohorsky1994}.

The scattering integral can be written as
\begin{align*}
  \mathcal{B}_S[f] = \frac{1}{2} \int (\epsilon')^2 d\epsilon' d\mu'  \left\lbrace R^{in}_0(\epsilon,\epsilon') f(\epsilon',\mu') [ 1-f(\epsilon,\mu) ] \right.  \\ \left. -  R^{out}_0(\epsilon,\epsilon') f(\epsilon,\mu) [ 1-f(\epsilon',\mu')] \right\rbrace \\
   \numberthis{}
\end{align*}
where $R^{in}_0$ and $R^{out}_0$ are the zeroth Legendre moment of the ingoing and outgoing scattering kernels:
\begin{equation}
    R^{in/out}_0(\epsilon,\epsilon') = \int_0^\pi R^{in/out}(\epsilon,\epsilon',\cos\Theta) \sin\Theta d\Theta~.
\end{equation}
They fulfil the in/out symmetry $R^{in}_0(\epsilon,\epsilon') = R^{out}_0(\epsilon',\epsilon)$, as well as the detailed balance condition $R^{in}_0(\epsilon,\epsilon') = \mathrm{e}^{(\epsilon' - \epsilon)/(k_B T)} R^{out}_0(\epsilon,\epsilon')$.

The effective transport coefficients can then be written as
\begin{align}
  j_{\mathrm{eff},S}(\epsilon) &= \frac{1}{2} \int (\epsilon')^2 d\epsilon' R^{in}_0 (\epsilon,\epsilon') J(\epsilon') \\
  \chi_{\mathrm{eff},S}(\epsilon) &= j_{\mathrm{eff},S}(\epsilon) + \frac{1}{2} \int (\epsilon')^2 d\epsilon' R^{out}_0(\epsilon,\epsilon') [1-J(\epsilon')]
\end{align}
where $J(\epsilon) = \frac{1}{2}\int f(\epsilon,\mu)d\mu$ is the zeroth moment of the distribution function.

\subsection{Pair production integral}

The dominant processes for neutrino pair production in dense nuclear
matter are mainly the nucleon-nucleon bremsstrahlung $N+N
\leftrightarrows N+N+\nu+\bar\nu$, and on a less important level the
electron-positron annihilation $e^- + e^+ \leftrightarrows \nu +
\bar\nu$. Bremsstrahlung is treated with the analytic fit of
\citet{Hannestad1997}, and electron-positron annihilation is treated in
the ultrarelativistic limit \citep[see][]{Bruenn1985}. The
ultrarelativistic approximation is not justified when the
PNS becomes cold enough such that the mean energy of
neutrinos gets close to the electron mass $m_e c^2$. But as this
process is subdominant a complete treatment should not generate
important changes in the results. Instead we simply add a cutoff to
emulate the $2m_e c^2$ threshold.

The pair production integral depends upon the distribution function of the antineutrino $\bar f$ :
\begin{align*}
  \mathcal{B}_{P}[f,\bar f] = \frac{1}{2} \int (\epsilon')^2 d\epsilon' d\mu'  \left\lbrace R^{p}_0(\epsilon,\epsilon') [1-f(\epsilon,\mu)] [ 1-\bar f(\epsilon',\mu') ] \right.  \\ \left. -  R^{a}_0(\epsilon,\epsilon') f(\epsilon,\mu) \bar f(\epsilon',\mu') \right\rbrace \\
   \numberthis{}
\end{align*}
where $R^{p}_0$ and $R^{a}_0$ are the zeroth Legendre moment of the production and absorption kernels for pair production,
\begin{equation}
    R^{p/a}_0(\epsilon,\epsilon') = \int_0^\pi R^{p/a}(\epsilon,\epsilon',\cos\Theta) \sin\Theta d\Theta~.
\end{equation}
They fulfil the detailed balance condition $R^a_0(\epsilon,\epsilon') = \mathrm{e}^{(\epsilon+\epsilon')/(k_B T)} R^p_0(\epsilon,\epsilon') $. 
The effective transport coefficients can then be written as
\begin{align}
  j_{\mathrm{eff},P}(\epsilon) &= \frac{1}{2} \int (\epsilon')^2 d\epsilon' R^{p}_0 (\epsilon,\epsilon') [1-\bar J(\epsilon') ] \\
  \chi_{\mathrm{eff},P}(\epsilon) &= j_{\mathrm{eff},P}(\epsilon) + \frac{1}{2} \int (\epsilon')^2 d\epsilon' R^{a}_0(\epsilon,\epsilon') \bar J(\epsilon')
\end{align}
where $\bar J(\epsilon) = \frac{1}{2}\int \bar f(\epsilon,\mu)d\mu$ is
the zeroth moment of the antineutrino distribution function.

\section{Phenomenological approach to include modified URCA processes}\label{app:murca}

The neutrino emissivity from charged current processes can be written
as \citep[see e.g.][for details]{Oertel2020}
\begin{align}
  j(E_\nu) =& - \frac{G_F^2 V_{ud}^2}{8} \int \frac{d^3 k_e}{ (2 \pi)^3}
              \frac{1}{E_e E_\nu} L^{\lambda\sigma} \mathrm{Im}
              \Pi^{R}_{\lambda\sigma}(q) \times \nonumber \\
            & f_F (E_e - \mu_e) (1 + f_B (q_0))
              +\mathrm{positronic\ contribution}
\label{eq:nuemissivity}
\end{align}
with a similar expression for antineutrinos and the mean free path
related via detailed balance. $G_F$ denotes here the Fermi coupling
constant and $V_{ud}$ the quark mixing matrix element entering the
charged current processes with nucleons.
$q = (E_e - E_\nu - \mu_e + \mu_\nu, \vec{k_e} - \vec{k_{\nu}})$ with
subscripts indicating electron or neutrino energy and chemical
potential, respectively, and $f_{F/B}$ are fermionic or bosonic
distributions functions. Assuming non-relativistic nucleons and
neglecting the momentum dependence of the nucleonic form factors, the
product of the lepton tensor $L^{\ls}$ and the retarded polarisation
$\Pi^R$ becomes
\begin{align}
  L^{\ls} \Pi^R_{\ls} &= 8 \left(\Pi_V ( 2 E_e E_\nu - K_e \cdot K_\nu)\right. \nonumber \\ &\qquad{} \left. + \Pi_A ( 2 E_e E_\nu + K_e\cdot K_\nu)\right)~, 
\end{align}
with a vector $\Pi_V$ and an axial part $\Pi_A$. In mean field
approximation for the direct processes, we obtain
\begin{equation}
  \mathrm{Im}\, \Pi_V(q)  = \mathrm{Im}\,\Pi_A(q) = 2 \mathrm{Im} \, L(q)~,
\label{eq:lindhardfull}
\end{equation}
with the well-known Lindhard function $L(q)$,
\begin{equation}
L(q) = \lim_{\eta \to 0} \int \frac{d^3 k}{( 2 \pi)^3}\frac{f_{F}(\ep - \mu_p^*) - f_F(\en-\mu_n^*)}{\tilde{q_0} + i \eta  + \ep - \en}~.
\label{eq:lindhard}
\end{equation}
with $\tilde{q_0} = q_0 + \mu^*_n - \mu^*_p$. The single particles
energies are $\epsilon_k^i = \frac{\vec{k}^2}{2 \, m_i^*} + m_i^*$
with effective masses $m_i^*$, and $\mu^*_{n/p}$ denote proton and
neutron effective chemical potentials. Values for the effective masses
and chemical potentials for the conditions of Figs.~\ref{fig:murcanu}
and \ref{fig:murcanubar} are listed in Table~\ref{tab:uint}. The
approach by \cite{Roberts2012neutrinos} to describe modified URCA
processes consists in considering a finite lifetime $\tau$ for the
quasi-particles entering the Lindhard function in the axial channel
leading to
\begin{align}
\Pi_{A}(q) &= 2 \int \frac{d^3 k}{( 2 \pi)^3}\frac{f_{F}(\ep - \mu_p^*) - f_F(\en - \mu^*_n)}{\tilde{q_0} + i \tau + \ep - \en} \nonumber \\ & \qquad \times \left(1-\frac{i}{\tau} \frac{1}{\en-\ep}\right)
\label{eq:lindhardgen}
\end{align}
The values for $\tau$ have been taken from \citet{Bacca2012}.

\end{appendix}


\bsp	
\label{lastpage}
\end{document}